\documentclass[nofootinbib,twocolumn,eqsecnum,floats,aps]{revtex4}
\usepackage{graphicx}
\usepackage{epsfig}
\usepackage{bm}
\usepackage{ulem,color,graphics}
\usepackage{multirow}
\usepackage{blindtext}
\usepackage[thicklines]{cancel}

\usepackage{xcolor}

\def\bit{\begin{itemize}}
\def\eit{\end{itemize}}
\def\bnu{\begin{enumerate}}
\def\enu{\end{enumerate}}
\def\sss{\scriptscriptstyle}

\def\A {{{\cal A}}}

\def\O {{{\cal O}}}

\def\nn{\nonumber }

\def\M {{{\cal M}}}
\def\x{\times}
\def\Ket#1{||#1 \rangle}
\def\Bra#1{\langle #1||}

\def\nn{\nonumber }
\def\be{\begin{equation}}
\def\ee{\end{equation}}
\def\br{\begin{eqnarray}}
\def\er{\end{eqnarray}}
\def\brn{\begin{eqnarray*}}
\def\ern{\end{eqnarray*}}

\def\bra#1{\langle #1|}
\def\ket#1{|#1 \rangle}
\def\rf#1{{(\ref{#1})}}
\def\ov#1#2{\langle #1 | #2  \rangle }
\def\sixj#1#2#3#4#5#6{\left\{\negthinspace\begin{array}{ccc}
#1&#2&#3\\#4&#5&#6\end{array}\right\}}
\def\ninj#1#2#3#4#5#6#7#8#9{\left\{\negthinspace\begin{array}{ccc}
#1&#2&#3\\#4&#5&#6\\#7&#8&#9\end{array}\right\}}
\def\go{\rightarrow  }
\def\etal {{\it et al.}}
\def\E {{{\cal E}}}

\def\fot{\frac{1}{2}}

\def\J {{{\cal J}}}

\def\a {{\alpha}}
\def\b {{\beta}}

\def\sss{\scriptscriptstyle}

\def\d{\dagger}
\def\w {\omega}

\begin{document}
\title{ A  Nuclear Structure Model for  Double Charge-Exchange Processes}
\author{ V. dos S. Ferreira$^{1}$,  A. R. Samana$^{2}$,
F. Krmpoti\'c$^{3}$, and M. Chiapparini$^{1}$ }
\affiliation{$^1$Instituto de F\'isica,
Universidade do Estado do Rio de Janeiro,
CEP 20550-900, Rio de Janeiro-RJ, Brazil}%
\affiliation{$^2$Departamento de Ci\^encias Exactas e Tecnol\'ogicas,
Universidade Estadual  de Santa Cruz,
CEP 45662-000 Ilh\'eus, Bahia-BA, Brazil}%
\affiliation{$^3$Instituto de F\'isica La Plata, CONICET,
Universidad Nacional de La Plata, 1900 La Plata, Argentina.}%
\begin{abstract}

A new model, based on the BCS approach, is specially designed to describe
nuclear phenomena  $(A,Z)\rightarrow (A,Z\pm 2)$ of
double-charge exchange (DCE). After being proposed, and applied in the
particle-hole limit,  by one of the authors (F.~Krmpoti\'c~\cite{Krm05}),
so far it was never been applied within the BCS mean-field framework,
nor has its ability to describe DCE processes been thoroughly explored.
It is a natural extension of the pn-QRPA model,  developed
by Halbleib and Sorensen~~\cite{Hal67} to describe the single $\b$-decays
$(A,Z)\rightarrow (A,Z\pm 1)$,  to the DCE processes.
As such, it  exhibits several  advantages over the  pn-QRPA model when
is used in the evaluation of  the double beta decay (DBD) rates. For
instance,
i) the extreme sensitivity of the nuclear matrix elements (NMEs) on the
model parametrization does not occur,
ii) it allows to study NMEs, not only for the fundamental state in
daughter nuclei, as the pn-QRPA model does, but also for all
final $0^+$ and $2^+$ states,  accounting at the
same time their excitation energies and the corresponding DBD
Q-values, iii) together with the  DBD-NMEs it provides also
the energy spectra  of  Fermi and Gamow-Teller DCE transition
strengths, as well as the locations of the corresponding
resonances and their   sum rules, iv) the latter are relevant
for both the DBD and the DCE reactions,  since the
involved nuclear structure is the same;  this correlation
does not exist within the  pn-QRPA model.
As an example, detailed  numerical calculations are presented
for the   $(A,Z)\rightarrow (A,Z+ 2)$ process
in $^{48}$Ca $\rightarrow ^{48}$Ti and
the $(A,Z)\rightarrow (A,Z- 2)$ process
in $^{96}$Ru $\rightarrow ^{96}$Mo,
involving all final $0^+$ states and $2^+$ states.

\end{abstract}


\maketitle

\section {Introduction}\label{Sec1}
The  Double Charge-Exchange  (DCE) processes relate the  $(A,Z)$ nuclei
with the  $(A,Z+2)$ and $(A,Z-2)$ nuclei
and will be labeled as $\{+2\}$ and  $\{-2\}$ processes respectively.

The most studied DCE process is the Double Beta Decay (DBD).
It is the slowest physical process observed so far,
and  can be used to learn about  neutrino physics,
provided we know how to deal with the nuclear structure.
According with the number and type of leptons we may have the following
DBD modes:
i)  double-electron decay ($2\b^-$),  ii) double-positron decay,
iii) electron capture-positron emitting decay ($e\b^+$),
and vi) double electron capture decay ($ee$). Each of these
decays occurs either with the emission of two neutrinos ($2\nu$-decay)
or they are neutrinoless ($0\nu$-decay). To simplify the notation and
when it does not cause confusion, we will designate the first process
as DBD$^-$ and the remaining three as DBD$^+$.

The $0\nu$-decay rates depend on several unknown parameters such as
neutrino mass, Majoron coupling, the coupling constants of the right-handed
components of the weak hamiltonian, etc.) and the only way to put these
in evidence is by having sufficient command over the nuclear structure.
It is precisely at this point that the  $2\nu 2\b^\mp$, $2\nu e\b^+$,
and $2\nu ee$  decay modes are important. A comparison between experiment
and theory for them provides a measure of the confidence that one may have
in the nuclear wave functions employed for extracting the unknown parameters
from $0\nu$-lifetime measurements.

The number of possible candidates for $2\beta^-$-decay is quite large:
there are 35 nuclei. In addition, 34 nuclei can undergo $2e$-electron capture,
while 22 and 6 nuclei can undergo $e\b^+$ and $2\b^+$-decays respectively~\cite{Bar11}.
The discovery of the massiveness of the neutrino, through the observation
of oscillations, boosted the importance of the $0\nu$-DBD, since they are
the only observables capable of providing the magnitude of the effective neutrino mass.

It is well known that the involved nuclear structure in $2\nu$
and $0\nu$-DBD is the same one that describes the DCE reactions.
This fact reignited recently the interest in the measurements of
heavy-ion-induced DCE reactions, such as the NUMEN project~\cite{Cap15,Cap18}
involving the $^{40}$Ca($^{18}$O,$^{18}$Ne)$^{40}$Ar process.
Simultaneously, the interest in the theoretical study of the DCE reactions
has been renewed~\cite{Sag16,Aue18}.
More, Shimizu,  Men\'endez, and Yako~\cite{Shi18} have also latterly noted
correlations between the DCE and $0\nu2\beta^-$-decay.
In all the mentioned theoretical studies, the calculations were made
within the framework of the shell model (SM).

The neutrinoless DBDs occur in medium-mass nuclei that are often far
from closed shells and, as a consequence, the calculations are mostly made
in the proton-neutron Quasiparticle Random Phase Approximation (pn-QRPA),
since this tool is computationally much more simple  than the SM. As discussed
in Ref.~\cite{Esc10}, the kind of correlations that these two methods include are not the same.
The pn-QRPA deals with a large fraction of nucleons in a large single-particle space, but within a
modest configuration space. The SM, by contrast, deals with a small fraction of nucleons
in a limited single-particle space, but allows them to correlate in arbitrary ways
within a large configuration space.
\begin{figure}[h]
\centering
\includegraphics[scale=0.50]{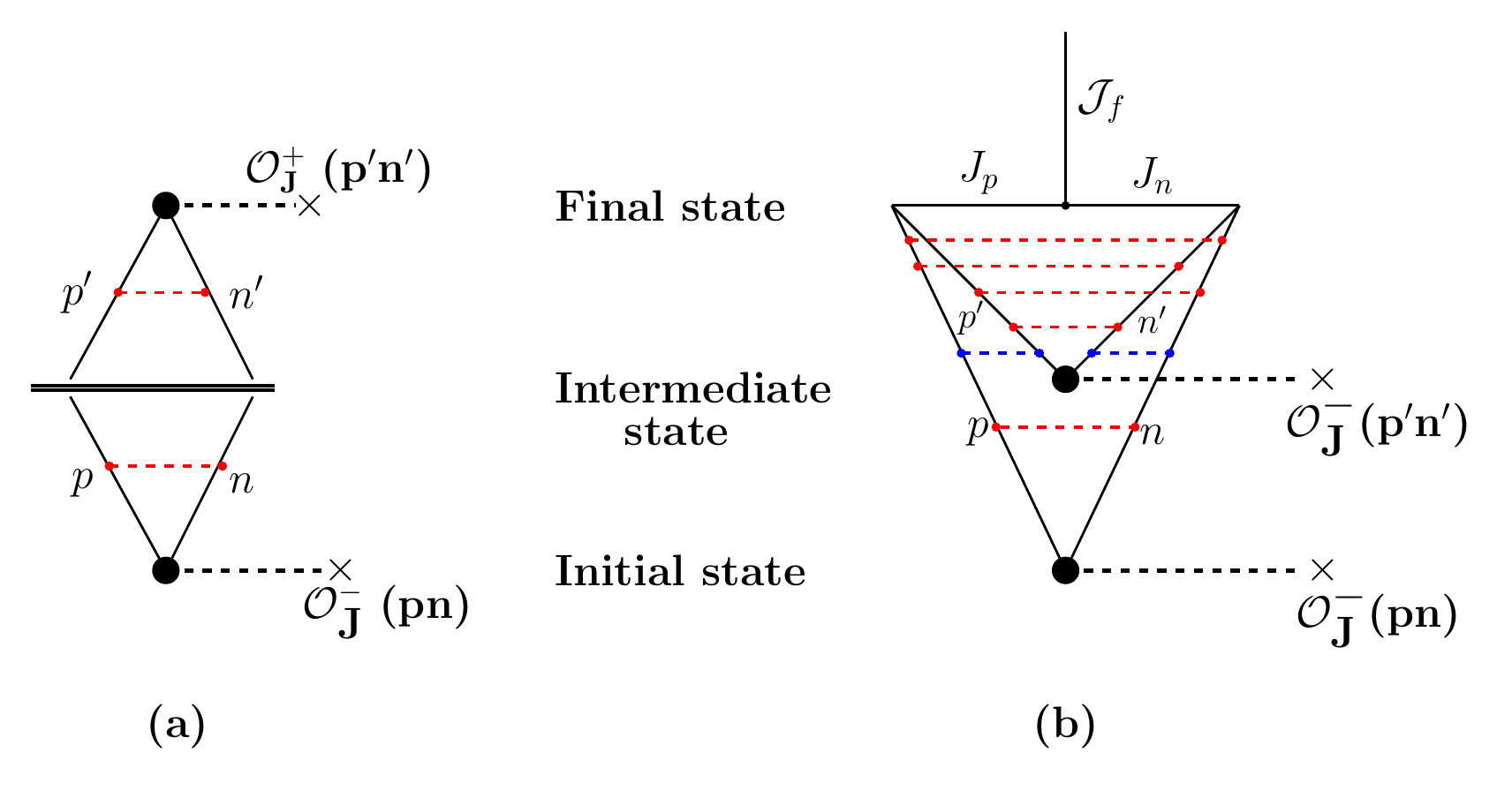}
\caption{\label{F1}(Color online)
Graphical representation of the numerators in the  $2\b^-$ NME, where
the black points indicate the single $\b$-decays. They are:
a) $\Bra{0^+_f}\O^+_J\Ket{J_\a^+}\ov{J_\a^+}{J_{\a'}^+}\Bra{J_{\a'}^+}\O^-_J\ket{0_i^{+}}$
in the pn-QRPA model,  where the overlap between the initial and final QRPA solutions
for the intermediate nucleus, $\ov{J_\a^+}{J_{\a'}^+}$, is represented by a thick line,
and b) $\Bra{\J^+_f}\O^-_J\Ket{J_\a^+}\Bra{J_\a^+}\O^-_J\ket{0_i^{+}}$ in the (pn,2p2n)-QTDA
model, which appears in Eqs. \rf{9} and \rf{15}, and indicates that
the first $\b^-$-decay is switched on {in the initial state and the second
in the intermediate state.}  The $2\b^+$NMEs  are represented
in the same way after making the substitution $\O^\mp_J\leftrightarrow \O^\pm_J$.
The pn, and nn + pp nuclear interactions between protons and neutrons are indicated
by red and blue dashed lines, respectively.
The five vertices of the diagram b) correspond to five of six  angular momentum
coupling  in the symbol 9j in Eq.~\rf{47}. The sixth coupling $(JJ)\J$
corresponds to the three unconnected  lines in this figure.}
\end{figure}
There is another important difference. The standard pn-QRPA only allows us to calculate
the double-charge exchange transitions from the ground state of the decaying $(A,Z)$ nucleus
to the ground state in the final $(A,Z\pm 2)$ nuclei. In fact, to evaluate the transitions
going to the excited states, a second (pp + nn)-QRPA must to be performed~\cite{Suh12,Del17},
which introduces additional free parameters and it is limited to one and two
quadrupole phonon states.

To deal with the DCE  processes we will resort here to a Tamm-Dancoff
Approximation (TDA),  which has been suggested  and discussed on its $ph$-limit
for  $^{48}$Ca  more than a decade ago in Ref.~\cite{Krm05}.
In this model is assumed that the initial, intermediate and final nuclei
are the BCS vacuum,  pn-excitations, and 2p2n-excitations respectively.
The resulting model will be labeled as (pn,2p2n)-QTDA. The main differences
with the standard pn-QRPA model are illustrated in Fig.~\ref{F1}.

The present model is a natural extension to double charge-exchange
processes of the pn-QRPA model, originally  proposed by Halbleib and Sorensen (HS)
in 1967 to describe the single $\beta$-decays~\cite{Hal67}.
As such, it allows to evaluate the NME, not only for  the ground states but
also of all final $0^+$ and $2^+$ states, as well as the $Q$-values for
the $2\beta^-$-decay ($Q_{2\b^-}$), and for the $2e$-capture ($Q_{2e}$).
It yields as  well the DCE energy  strength spectra  and  their sum
rules, which are relevant for associated reaction processes and resonances.
Detailed numerical calculations are performed in the
present work for the $^{48}$Ca $\go ^{48}$Ti,  and $^{96}$Ru $\go ^{96}$Mo
processes, involving their final $0^+$ and $2^+$ states.

\section {Formalism}\label{Sec2}

\subsection{Nuclear matrix elements and double charge-exchange excitations}

Independently of the nuclear model used, and when are only considered the allowed
transitions, {\it i.e.} the Gamow-Teller (GT) and Fermi (F) transitions, the NMEs
for the $2\nu2\beta^{\pm}$-decay, from the ground state
$0^+_i$ in the initial nucleus $(A,Z)$ to one of the states $0^+_f$ in the
final nuclei $(A,Z\mp 2)$, reads~\footnote{For the first-forbidden NME see
Ref. \cite{Bar95}.}
\begin{eqnarray}
&& M^{2\nu^\pm}(0^+_f)=  M^{2\nu^\pm}_{F}(0^+_f)+  M^{2\nu^\pm}_{GT}(0^+_f)\label{1}
\\
&\equiv&
\sum_{J=0,1}(-)^Jg_{\sss J}^2
\sum_{\a} \left[\frac{\Bra{0^+_f}\O^{\pm}_J\Ket{J^{+}_\a}
\Bra{J^{+}_\a}\O^\pm_J\Ket{0^{+}_i}}{{\cal D}_{J_\a,0_f}^{2\nu^\pm}} \right],
\nn
\end{eqnarray}
where $g_{\sss 0}\equiv g_{\sss V}$ and $g_{\sss 1}\equiv g_{\sss A}$ are
the vector and axial-vector weak coupling constants respectively, and
the summation goes over all intermediate virtual states
$\mid J^{+}_\a\rangle$ in the  nuclei $(A,Z\mp1)$.

The one-body operators are
\br
\O^{-}_J&=&
\hat{J}^{-1}\sum_{pn}\Bra{p}{\rm O}_J\Ket{n}\left(c^{{\dagger}}_p c_{\bar{n}}\right)_J,
\nn\\
\O^{+}_J&=&
\hat{J}^{-1}\sum_{pn}\Bra{n}{\rm O}_J\Ket{p}\left(c^{{\dagger}}_n c_{\bar{p}}\right)_J,
\label{2}
\end{eqnarray}
with ${\rm O}_0=1 $, and ${\rm O}_1=\sigma$  for F and GT transitions, respectively,
and $c^\d_k\equiv c^\d_{j_k,m_k}$ and $ c_{\bar k}\equiv (-)^{j_k-m_k}c_{j_k,-m_k}$
being the single-particle creation and annihilation operators, and $\hat{J}=\sqrt{2J+1}$.

The energy denominator in \rf{1} is~
\footnote{The last term in the denominator \rf {3} is based on the assumption
that the lepton energies can be replaced by $e+ \nu \cong (E_{0^+}^{\{0\}}-E^{\{\mp 2\}}_{\J_f})/2$,
whose validity for the mixed mode was questioned by Hirsch et. al \cite{Hir94}.
Following their idea on equal sharing of the liberated energy among  the emitted
leptons, Suhonen~\cite{Suh12} has derived formulas for the three
different DBD$^+$, and has used them in the evaluation of  the decay
$^{96}$Ru $\rightarrow ^{96}$Mo. Unfortunately, the author has omitted a factor of 2 in his
Eq.(13), which makes wrong the numerical result of mode $2\b^+$.
The same error was propagated in subsequent works~\cite{Pir15,Li17}.
Here we will continue using the estimate~\rf{3}.}
\begin{eqnarray}
{\cal D}_{J_\a,\J_f}^{2\nu^\pm}
&=&E_{J_\a }^{\{\mp 1\}}-\frac{E_{0^+}^{\{0\}}+E^{\{\mp 2\}}_{\J_f}}{2}
\nn\\
&=&E_{J_\a }^{\{\mp 1\}}-E_{0^+}^{\{0\}}+\frac{E_{0^+}^{\{0\}}-E^{\{\mp 2\}}_{\J_f}}{2},
\label{3}\end{eqnarray}
where $E_{0^+}^{\{0\}}$, $E_{J_\a}^{\{\pm 1\}}$ and $E_{\J_f}^{\{\pm 2\}}$
are the energies of the decaying  $(A,Z)$ nucleus in its
ground state, the intermediate $(A,Z\pm 1)$ nuclei in the state $J_\a ^{+}$,
and final $(A,Z\pm 2)$  nuclei in the state $\J_f^{+}$, respectively.

The matrix elements $\Bra{J^{+}_\a}\O^\pm_J\Ket{0^{+}_i}$  can be expressed as
a product of model dependent one-body densities
\br
\rho^{-}(pnJ_\a)&=&
\hat{J}^{-1}\Bra{{J_\a}}(c^{{\dagger}}_{p}c_{\bar{n}})_{J}\Ket{0^+_ i},
\nn\\
\rho^{+}(pnJ_\a) &=&
\hat{J}^{-1} \Bra{J_\a}(c^{{\dagger}}_{n}c_{\bar{p}})_{J}\Ket{0^+_i},
\label{4} \er
and a purely geometric (angular) factors~
\footnote{We use here the angular
momentum coupling scheme $\ket{({1 \over 2},l)j}$.}
\br
&& O_J(pn)\equiv\Bra{p}{\rm O}_J\Ket{n}
\nn\\
&=&\delta_{l_pl_n}\sqrt{2} \hat{J}\hat{j}_n\hat{j}_p
(-)^{j_p+l_p+J+\fot}\sixj{j_p}{j_n}{J}{\fot}{\fot}{l_n},
\label{5}\er
which is the single-particle NME, with $O_J(np)=(-)^{j_n-j_p}O_J(pn)$.
Namely,
\br
\Bra{J^+_\a}\O^\pm_J\Ket{0^{+}_i}=\sum_{pn}\rho^\pm(pnJ_\a^\pi) O_J(pn).
\label{6} \er
Similarly,
\br
\Bra{\J^+_f}\O^\pm_J\Ket{J^+_\a}=\sum_{np}\rho^\pm(pnJ_\a,\J^+_f) O_J(pn),
\label{7} \er
where
\br
\rho^{+}(pnJ_\a,\J^+_f)&=&
\hat{J}^{-1} \Bra{\J^+_f}(c^{{\dagger}}_{n}c_{\bar{p}})_{J}\Ket{{J_\a}},
\nn\\
\rho^{-}(pnJ_\a,\J^+_f)&=&
\hat{J}^{-1} \Bra{\J^+_f}(c^{{\dagger}}_{p}c_{\bar{n}})_{J}\Ket{{J_\a}},
\label{8} \er
are the  corresponding density matrices for transitions from the
intermediate states $\ket{J_\a}$ to the final states $\ket{\J^+_f}$,
with $\J=0,2$.

The corresponding matrix elements are:
\br
&& M^{2\nu^\pm}_{F}(0^+_f)=g_{\sss V}^2
\sum_{\a }\frac{\Bra{0^+_f}\O^\pm_0\Ket{0_\a^+}
\Bra{0_\a^+}\O^\pm_0\ket{0_i^+}}{{\cal D}_{0^+_\a,0^+_f}^{2\nu^\pm}},
\nn\\
 &&M^{2\nu^\pm}_{GT}(\J^+_f)=\frac{-g_{\sss A}^2}{\sqrt{\J+1}}
\sum_{\a }\frac{\Bra{\J^+_f}\O^\pm_1\Ket{1_\a^+}
\Bra{1_\a^+}\O^\pm_1\ket{0_i^+}}
 {\left({\cal D}_{1^+_\a,\J^+_f}^{2\nu^\pm}\right)^{\J+1}},
\nn\\
\label{9}\er
where  the GT-NMEs to  $2^+_f$ states have also been
included~\cite{Doi92,Rad07,Shu09}.

All the information about the nuclear structure is contained in
the one-body density matrices \rf{4} and \rf{8}, or more precisely in
the two-body density matrices
\begin{eqnarray}
\rho^\pm(pnp'n';J_\a,\J^+_f)
&=&\rho^{\pm}(pn;J_\a)\rho^{\pm}(p'n';J_\a,\J^+_f).
\nn\\
\label{10}\end{eqnarray}

The NME for the  $0\nu2\beta^\pm$-decays to the $0^+_f$ final states
can be easily evaluated from these densities.  In fact, after
doing  in~\cite[Eqs. (2.20)]{Fer17} the replacement
\begin{eqnarray}
\rho^{ph}(pnp'n';J_\a)\go \rho^\pm(pnp'n';J_\a,0^+_f),
\label{11}\end{eqnarray}
 we can express them as
\br
{ M}^{0\nu^\pm}(0^+_f)=\sum_X{ M}^{0\nu^\pm}_{X}(0^+_f),
\label{12}
\end{eqnarray}
where $X=V, A,P, M$ stands for vector ($V$), axial-vector ($A$),
pseudo-scalar ($P$), and weak-magnetism ($M$) terms.
We proceed in the same way with the NME ${ M}^{0\nu^\pm}(2^+_f)$.

It is well known that the single $\b$-decay processes to
the states $J_\a$ in $(A,Z+1)$ and $(A,Z-1)$ nuclei are
related to the following single charge-exchange transition strengths
\br
 S^{\{\pm 1\}}_J&\equiv&\sum_{ \a}B^{\{\pm 1\}}_{J_{\a}}=\hat{J}^{-2}
\sum_{\a}|\Bra{J_\a^+}{\O^\mp_J}\Ket{0^+_i}|^2.
\nn\\
\label{13}\er
When $\ket{J_\a^+}$ is a complete set of excited states
that can be reached by operating with $\O^\pm_J$ on the initial
state $\ket{0^+_i}$, they satisfy the single-charge
exchange  (SCE) sume rule or Ikeda sum rule, for both the F
and  GT transitions,
\br
&& S^{\{ 1\}}_J\equiv  S^{{\{+ 1\}}}_J- S^{{\{- 1\}}}_J
\nn\\
&=&(-)^J
\hat{J}^{-2}\bra{0^+_i}[\O^+_J,
\O^-_J]_0\ket{0^+_i}=N-Z.
\label{14}\er

Similarly, both ${ M}^{2\nu2\beta^\pm}(\J^+_f)$ and
${ M}^{0\nu2\beta^\pm}(\J^+_f)$ are related to the
double-charge-exchange operators $(\O^\pm_J\O^\pm_J)_\J$ and
to their spectral distributions in nuclei $(A,Z \pm 2)$ nuclei given by
\br
 S^{\{\pm 2\}}_{J\J}&\equiv&\sum_{ f}B^{\{\pm 2\}}_{J\J_f}
\nn\\
&=&\hat{J}^{-2}\sum_{ f}
|\sum_{\a }\Bra{\J^+_f}\O^{\mp}_J\Ket{J_\a^+}
\Bra{J_\a^+}\O^{\mp}_J\ket{0_i^{+}}|^2.
\nn\\
\label{15}\end{eqnarray}
When both $\ket{J_\a^+}$ and $\ket{\J_f^+}$ are complete set of excited states
that can be reached by operating with $\O^\pm_J$, and  $(\O^\pm_J\O^\pm_J)_\J$
on the initial state $\ket{0^+_i}$, their differences
\br
 S^{\{ 2\}}_{J\J}&=&  S^{{\{+ 2\}}}_{J\J}- S^{{\{- 2\}}}_{J\J}
\nn\\
&=&\hat{J}^{-2}\sum_{ f}\left[
\left|\sum_{\a }\Bra{\J^+_f}\O^-_J\Ket{J_\a^+}
\Bra{J_\a^+}\O^-_J\ket{0_i^{+}}\right|^2
\right.
\nn\\
&-&\left.\left|\sum_{\a }\Bra{\J^+_f}\O^+_J\Ket{J_\a^+}
\Bra{J_\a^+}\O^+_J\ket{0_i^{+}}\right|^2
\right]\label{16}\end{eqnarray}
obey the double-charge-exchange sum rules (DCESR), which
were evaluated in Refs. \cite{Aue18,Vog88,Mut92,Zhe89,Sag16}
with  the following results:
\begin{eqnarray}
{\sf S}_{DF} &\equiv& {\sf S}^{\{ 2\}}_{00} =2(N-Z)(N-Z-1), \label{17}\\
{\sf S}_{DGT,0}&\equiv&  {\sf S}^{\{ 2\}}_{10} \label{18} \\
&=&2(N-Z)\left(N-Z+1+2S^{\{+1\}}_1\right)
-\frac{2}{3}C\nn, \\
{\sf S}_{DGT,2}&\equiv&  {\sf S}^{\{ 2\}}_{12}\label{19} \\
&=&10(N-Z)\left(N-Z-2+2S^{\{+1\}}_1\right)+\frac{5}{3}C,\nn
\end{eqnarray}
where $C$ is a relatively small quantity given by~\cite[Eq. (4)]{Mut92}.
These equations agree with the Eq.~(8) in Ref.~\cite{Aue18} except
for a factor of 3 and the omission of $S^{\{+1\}}_1$.

Combining Eqs.~\rf{18} and \rf{19} one obtains the sum rule
for the total GT strength
\br
 {\sf S}_{DGT}
&=&12(N-Z)\left(N-Z-\frac{3}{2}+2S^{\{+1\}}_1\right)+C,
\nn\\
\label{20}\end{eqnarray}
which is independent of the structure of the ground-state
wave function~\cite{Mut92}.

The relationship between the $\b\b$-decay and double charge exchange
reactions has been discussed recently in Refs.~\cite{Cap15, Shi18}.

\subsection{(pn,2p2n)-QTDA Model}

The pn-QRPA evaluations of the $\b\b$-decays are generally limited
to the ground state of the final nuclei, {\it i.e.} to the calculation of
${ M}_{2{\nu}}(0^+_1)$ and ${ M}_{0{\nu}}(0^+_1)$. Moreover,
the SCE sume rules~\rf{14} are fulfilled within this model, but it does not allow
us to evaluate the strengths $ S^{\beta\beta^{\pm}}_{J\J}$ given by \rf{16}
and to discuss the corresponding DCESR listed in  Eqs. \rf{17}-\rf{20}.

To describe the intermediate states $J_\a^+$ in both pn-QRPA and
(pn,2p2n)-QTDA models it is used a nuclear Hamiltonian of the type
\be
H_1=H_0+H_{pn},
\label{21}\ee
where
\be
H_0=\sum_\a E_k a_k^\dag a_k
\label{22}\ee
is the independent-quasiparticle Hamiltonian, with
$a^\d_\a$ and $ a_{\bar{\a}}$ being the single-quasiparticle creation
and annihilation operators, defined by the Bogoljubov
transformation~\cite[Eqs.~(13.10)]{Suh07}
\br
a^\d_k&=&u_ac^\d_k+v_k c_{\bar{k}},
\nn\\
a_{\bar{k}}&=&u_a c_{\bar{k}}-v_ac^\d_k.
\label{23}\er
The proton and neutron pairing interactions are contained in
the transformation coefficients $u_a$ and $v_a$, and in the
quasiparticle energy
\be
E_k=[(e_k-\lambda)^2+\Delta_k^2]^\fot,
\label{24}\ee
where $e_k$ is the shell-model single-particle energy (spe),
and $\lambda$ is the chemical potential or Fermi level.
The energy gap parameters $\Delta_k$ and the pairing coupling constants
are determined  to reproduce the experimental odd-mass difference for
each nucleus. Finally, $H_{pn}$ is the quasiproton-quasineutron interaction.

Within the pn-QRPA, the states $\ket{J_\a}$  with excitation energy $\omega_{J_\a}$,
are created from the correlated initial and final $0^+$ ground states by
proton-neutron phonon creation operators $Q_{J_\a}$,
which are defined as a linear superposition of creation and annihilation
proton-neutron quasiparticle pair operators
\be
A^\dag(pnJ)=[a^\dag_pa^\dag_n]_{J},
\label{25}\ee
that is
\br
\ket{J_\a}&=&Q^\dag_{J_\a}\ket{0^+}
\label{26}
\\&\equiv&\sum_{pn}\left[ X^\a_{pnJ}A^\dag(pnJ)-Y^\a_{pnJ}A(pnJ)\right]\ket{0^+},
\nn
\er
and
\br
&&Q_{J_\a}\ket{0^+}=0, \hspace{.5cm} H_1\ket{J_\a}=\omega_{J_\a}\ket{J_\a}.
\label{27}\er

Usually this is done for initial $\ket{0^+_I}$ and final $\ket{0^+_F}$ ground states,
obtaining two sets of intermediate states $J_\a$  and $J_{\a'}$ in the $(N-1,Z+1)$
nucleus, which are different from each other.
Therefore, in the evaluation of the $\b\b$-NME it is necessary to consider their
overlap, which is indicated in Fig.~\ref{F1}, which corresponds to the substitution
\br
\sum_{J_\a}
&&\rho^{ph}(pnp'n';J_\a)\go\ov{0^+_I}{0^+_F}\x
\nn\\
&&\sum_{J^\pi \a\a'}\rho^{+}(p'n';{ J}_{\a'})\ov{{ J}_{\a'}}{J_\a}\rho^{-}(pn;J_\a).
\label{28}\er

The ground state defined in \rf{27} is more accurate that the BCS ground
state ($a_k\ket{BCS}=0$) since it contains terms with $0, 4, 8, \cdots$
quasiparticles~\cite{Esc10}. Nevertheless, in the present model we approximate
the initial ground state in the $(A,Z)$ nucleus by the BCS vacuum and
the states $\ket{J_\a}$ in the intermediate $(A,Z\mp 1)$ nuclei as
\br
\ket{J_\a}&=&\sum_{pn} X_{pnJ_\a}A^\dag(pnJ)\ket{BCS},
\nn\\
H_1\ket{J_\a}&=&\omega_{J_\a}\ket{J_\a}.
\label{29}\er
This disadvantage of the present model is counteracted by the description
that we make of the final states in the $(A,Z \mp 2)$ nuclei. That is,
instead of the correlated $\ket{0^+_F}$ state defined in \rf{26}, we have
\br
\ket{\J^{+}_f}&=&\sum_{p_1p_2n_1n_2J_{n}J_{p}} Y_{{\rm p_1p_2J_{p},n_1n_2J_{n}};\J^{+}_f}
\nn\\
&\x&\ket{p_1p_2J_{p},n_1n_2J_{n};\J^{+}}_A,
\label{30}\er
where $\J^+=0^+, 2^+$  and
\br
&&\ket{p_1p_2J_{p},n_1n_2J_{n};\J^{+}}_A
\nn\\&=&\left[\A^\d(p_1p_2J_{p}) \A^\d(n_1n_2J_{n})\right]^{\J^{+} }
\ket{BCS},
\label{31}\er
are antisymmetrized and normalized two-proton-two-neutron quasiparticle
states, with
\br
\A^\d(abJ)&=&N(ab)A^\d(abJ), \hspace{.4cm}
\nn\\
N(ab)&=&\frac{1}{\sqrt{1+\delta_{ab}}},
\label{32}\er
being normalized two-quasiparticle states.

The amplitudes $Y_{{\rm p_1p_2J_{12},n_1n_2J'_{12}};\J^{+}_f}$
are obtained by diagonalizing the Hamiltonian
 \cite{Pal66,Raj69}
\br
&&H_2=H_0+H_{pn}+H_{nn}+H_{pp},
\nn\\
&&H_2\ket{\J^{+}_f}    =\w_{\J^{+}_f}\ket{\J^{+}_f},
 \label{33}\er
in the basis ~\rf{31} with $H_{nn}$ and $H_{pp}$ being neutron-neutron
and proton-proton interactions.
Details on the evaluation of matrix elements of $H_2$ can be found
in References~\cite{Pal66,Raj69}. However, our final results are
different.

The  matrix element of $H_{pn}$ for the odd-odd nucleus reads
\br
&&\bra{BCS}A(npJ)H_{pn}A^\d(n'p' J) \ket{BCS}
\nn\\
&=& G(npn'p' J)(u_{p}u_{n}u_{p'}u_{n'}+v_{p}v_{n}v_{p'}v_{n'})
\nn\\
&+& F(npn'p'J)(u_{p}v_{n}u_{p'}v_{n'}+v_{p}u_{n}v_{p'}u_{n'}),
\label{34}\er
where the functions $G$ and $F$ are defined in the standard way~\cite{Bar60}.

The matrix elements of $H_{pn}$ in the basis~\rf{31} are derived
by employing the relation (1A-25) from~\cite{Boh69}. We get
\begin{widetext}
\br
&&\bra{BCS}\left[\A^\d(n_1n_2J_{n}) \A^\d(p_1p_2J_{p})\right]^{\J\d } H_{pn}
\left[\A^\d(n'_1 n'_2 J_{n}') \A^\d(p'_1 p'_2J_{n}')\right]^\J \ket{BCS}
\nn\\
&&=\hat J_{n}\hat J_{p}\hat J'_{n}\hat J'_{p}
N(n_1n_2)N(n'_1 n'_2)N(p_1p_2)N(p'_1 p'_2)
{\bar P}(n_1n_2 J_{n}){\bar P}(p_1p_2J_{p})
\nn\\&\x&
{\bar P}(n'_1 n'_2J_{n}'){\bar P}(p'_1 p'_2J_{p}')
\sum_{J_{1}J_{2}}
\hat J^2_{1}\hat J^2_{2}\ninj{n_1}{n_2}{J_{n}}{p_1}{p_2}{J_{p}}{J_{1}}{J_{2}}
{\J}\ninj{n'_1}{n'_2}{J'_{n}}{p'_1}{p'_2}{J'_{p}}{J_{1}}{J_{2}}{\J}
\nn\\
&\x&\bra{BCS}\A^\d(n_1p_1J_{1}) H_{pn}\A^\d(n'_1 p'_1 J_{1})
\ket{BCS}\delta_{p_2p'_2}\delta_{n_2n'_2}
\label{35}, \er
\end{widetext}
where the operator
\br
{\bar P}(p_1p_2J)=1+(-)^{p_1-p_2+J}P(p_1\leftrightarrow p_2),
\label{36}\er
exchanges the particles $p_1$ and $p_2$.

Finally, the matrix element of the neutron-neutron
Hamiltonian $H_{nn}$ in the same basis is
\begin{widetext}
\br
&&\bra{BCS}\left[\A^\d(n_1n_2J_{n})
\A^\d(p_1p_2J_{p})\right]^{\J\d } H_{nn}
\left[\A^\d(n'_1 n'_2 J_{n}') \A^\d(p'_1 p'_2J_{p}')\right]^\J
\ket{BCS}
\nn\\
&=&\delta_{J_{p}J'_{p}}\delta_{J_{n}J'_{n}}\delta_{p_1p'_1}\delta_{p_2p'_2}
\bra{BCS}\A^\d(n_1n_2J_{n}) H_{nn}\A^\d(n'_1n'_2J_{n})\ket{BCS}\nn\\
&=&\delta_{J_{p}J'_{p}}\delta_{J_{n}J'_{n}}\delta_{p_1p'_1}\delta_{p_2p'_2}
N(n_1 n_2 )N(n'_1 n'_2 )
\nn\\
&\x&\left[\right.(u_{n_1} u_{n_2} u_{n'_1} u_{n'_2}+v_{n_1} v_{n_2} v_{n'_1} v_{n'_2})
G(n_1n_2n'_1n'_2J_{n})
\nn\\
&+& (u_{n_1} v_{n_2} u_{n'_1} v_{n'_2}+v_{n_1} u_{n_2} v_{n'_1} u_{n'_2})
F(n_1n_2n'_1n'_2J_{n})
\nn\\
&-& (-1)^{n_1+n_2-J_{12}}(u_{n_1} v_{n_2} v_{n'_1} u_{n'_2}
                   +v_{n_1} u_{n_2} u_{n'_1} v_{n'_2})
F(n_2n_1n'_1n'_2J_{n})
\left.\right],\label{37}\er
\end{widetext}
and analogously for the proton-proton Hamiltonian $H_{pp}$.

The energies in the  denominator  ${{\cal D}_{J_\a,\J_f}^{2\nu^\pm}}$,
defined by~\rf{3}, are
\begin{eqnarray}
E_{J_\a }^{\{\pm 1\}}-E_{0^+}^{\{0\}}&=&
\w_{J_\a}\pm\lambda_p\mp\lambda_n,
\nn\\
E^{\{\pm 2\}}_{\J_f}-E_{0^+}^{\{0\}}&=&
\w_{\J_f}\pm2\lambda_p\mp 2\lambda_n,
\label{38}\end{eqnarray}
where $\lambda_p$  and $\lambda_n$ are the proton and neutron
chemical potentials. Therefore, for both  $2\b^+$ and
$2\b^-$-decays, they are
\br
{{\cal D}_{J_\a,\J_f}^{2\nu^\pm}}
\equiv{{\cal D}_{J_\a,\J_f}^{2\nu}}&=&\w_{J_\a}-\frac{\w_{\J_f}}{2}.
\label{39}\er
The lowest energies  $E^{\{\pm 2\}}_{0^+_f}$ are directly related
with the $Q$-values for the $2\beta^-$-decay ($Q_{2\b^-}$) and for the
$2e$-capture ($Q_{2e}$), defined as
\footnote{The $2\b^ +$ and $\b^ +e$ Q-values are:
\brn
Q_{2\b^+}&=&\M(Z,A)-\M(Z-2,A)-4m_e,
\nn\\
Q_{\b^+e}&=&\M(Z,A)-\M(Z-2,A)-2m_e.
\ern}
\br
Q_{2\b^-}&=&\M(Z,A)-\M(Z+2,A),
\nn\\
Q_{2e}&=&\M(Z,A)-\M(Z-2,A),
\label{40}\er
where the $\M$'s are the atomic masses. Namely,
\br
Q_{2\b^-}&=&E_{0^+}^{\{0\}}-E^{\{+2\}}_{0^+_1}
=-\w_{0_1^+}-2(\lambda_p- \lambda_n),
\nn\\
Q_{2e}&=&E_{0^+}^{\{0\}}-E^{\{-2\}}_{0^+_1}
=-\w_{0_1^+}+2(\lambda_p- \lambda_n).
\nn\\
\label{41}\er
Note that $Q_{2e}-Q_{2\b^-}=4(\lambda_p-\lambda_n)$,
and $Q_{2e}+Q_{2\b^-}=-2\w_{0_1^+}$.

To evaluate the one-body densities~\rf{4} and \rf{8}
we make use of \cite[Eqs. (15.4)]{Suh07} to get
\br
\left(c^{{\dagger}}_p c_{\bar{n}}\right)_{J}&\go u_nv_pA^\dag(npJ),
\nn\\
{\left(c^{{\dagger}}_{n} c_{\bar{p}}\right)_{J}}&\go u_pv_nA^\dag(pnJ),\label{42}\er
which from \rf{2} and \rf{5} immediately yields
\begin{eqnarray}
\rho^\pm(pnJ_\a)&=& X_{pnJ_\a} \left\{\begin{array}{ll}
u_{n}v_{p} \;\\
u_{p}v_{n} \;\\
\end{array}\right\},
\label{43}
\end{eqnarray}
and
\br
\Bra{J^+_\a}\O^\pm_J\Ket{0^{+}_i}&=&\sum_{pn} X_{pnJ_\a}O^\pm_J(pn) ,
\label{44}\er
with
\br
O^\pm_J(pn)&=&O_{J}(pn)\left\{\begin{array}{ll}
u_{n}v_{p} \;\\
u_{p}v_{n} \;\\
\end{array}\right\}.
\label{45} \er

The derivation of $\Bra{\J^+_f}\O^{\pm}_J\Ket{J^+_\a}$
is more laborious and one gets
\br
&&\Bra{\J^+_f}\O^{\pm}_J\Ket{J^+_\a}=
\hat{J}\hat \J_f
\sum_{pnp'n'J_{p}J_{n}} (-)^{J_p+J_n}\hat{ J_p}\hat{J_n}\nn\\
&\x&N(nn')N(pp') Y_{{\rm pp'J_{p},nn'J_{n}};\J^{+}_f}
{\bar P}(nn' J_n){\bar P}(pp'J_p)
\nn\\
&\x&
\ninj{p}{p'}{J_{p}}{n}{n'}{J_{n}}{J}{J}{\J} X_{p'n'J_\a}
O^\pm_J(pn).
\label{46}\er
The  densities $\rho^{\pm}(pn;J_\a^\pi,\J^+_f)$ result
immediately \rf{7} and \rf{46}.

Making use of orthogonality and completeness of both basis
$\ket{J_\a}$ and $A^\dag(pnJ)\ket{BCS}$ in \rf{29},
the relation \rf{15} can be expressed in a more compact form.
Namely as,
\br
 B^{\{\pm 2\}}_{J\J}
&=&\hat\J^{2}\left|\sum_{pp'nn'J_pJ_n} (-)^{J_p+J_n}
\hat{ J_p}\hat{J_n} N(nn')N(pp')\right.
\nn\\
&\x&\left.Y_{{\rm pp'J_{p},nn'J_{n}};\J^{+}_f}
{\bar P}(nn' J_n){\bar P}(pp'J_p)\right.
\nn\\
&\x&\left.
\ninj{p}{p'}{J_{p}}{n}{n'}{J_{n}}{J}{J}{\J}
O^\mp_J(p'n')O^\mp_J(pn)\right|^2.
\label{47}\end{eqnarray}
In this way, the transition strength turns out to
be independent of the intermediate states.

It is important to emphasize that the permutation operators
in the last two equations only act on the right side.
The physical meaning of these permutations  can be inferred
from the diagram (b) in Fig. \ref{F1},  where is  graphically represented
the DCE matrix element  $\sum_{\a }\Bra{\J^+_f}\O^-_J\Ket{J_\a^+}
\Bra{J_\a^+}\O^-_J\ket{0_i^{+}}$. This quantity is used in the evaluation
of both the $\b\b$-decay NME \rf{9} and the DCE transition strengths~\rf{15},
but the Eq.~\rf{47}  is applicable only in the latter case.

Together with the NME $M^{2\nu^\pm}(\J^+_f)$ given by \rf{1} and \rf{9} with
$M^{2\nu^\pm}(2^+_f)\equiv M_{GT}^{2\nu^\pm}(2^+_f)$, we will also evaluate
the half-lives $\tau_{2\nu^\pm}^{\a}(\J^+_f)$
for different  ${\a}(=2\b^-,2\b^+,  e\b^+,2e)$.
This is done from
\br
[\tau_{2\nu^\pm}^{\a}(\J^+_f)]^{-1}=
g_{\sss A}^4 \left|M^{2\nu^\pm}(\J^+_f)\right|^2G^\a_{2\nu}(\J^+_f),
\label{48}\er
{\it i.e.} the product of dimensionless axial vector coupling constant,
$g_{\sss A}$, common NME, $M^{2\nu^\pm}(\J^+_f)$, given in natural
units ($\hbar=m_e=c=1$), and different leptonic kinematics factors,
$G^\a_{2\nu}(\J^+_f)$, in yr$^{-1}$. The last ones can be found in ~\cite [Table II]{Kot13}
for several nuclei of interest.
(For  the most recent  computations of phase space factors see Refs.~\cite{Sto19,Sto19a}.)

The  excitation energies in the final nuclei are  calculated from
\be
\E_{f}=E^{\{+2\}}_{0^+_f}-E^{\{+2\}}_{0^+_1}.
\label{49}\ee
It should be noted that, just as in the pn-QRPA model the excitation
energies in the $(Z,A\pm 1)$ nuclei are the same, in the current model
the excitation energies in the $(Z,A\pm 2)$ nuclei are the same.

Finally, the centroid energies of the DCE transition strengths are
defined as
\be
{\bar E_{J\J}}^{\{\pm 2\}}=\frac{\sum_f\E_{\J^+_f}
B^{\{\pm 2\}}_{J\J_f}}{ S^{\{\pm 2\}}_{J\J}}.
\label{50}\ee
\begin{figure*}[t]
\begin{center}
\begin{tabular}{cc}
\includegraphics[width=8cm]{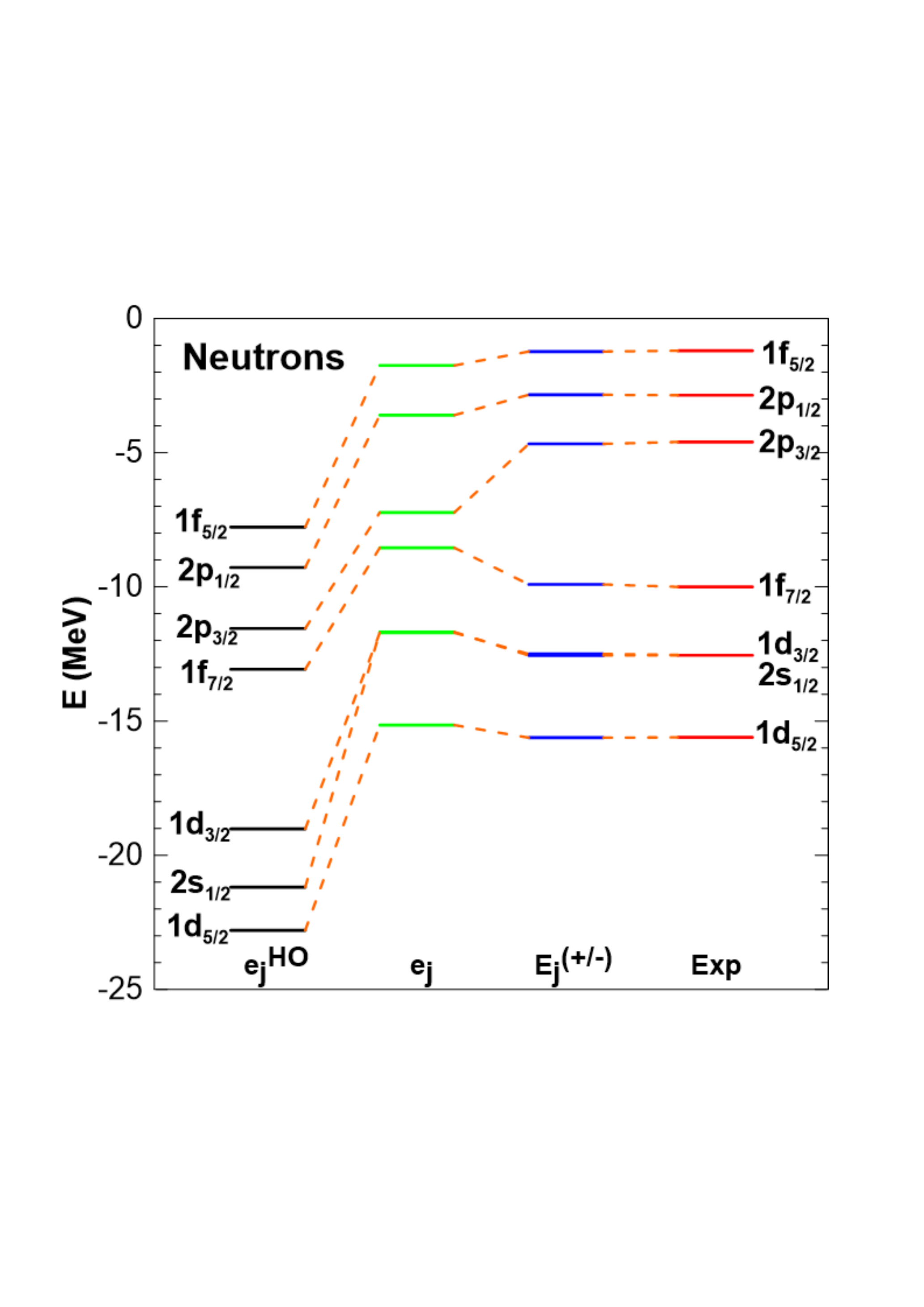}
&
\includegraphics[width=8cm]{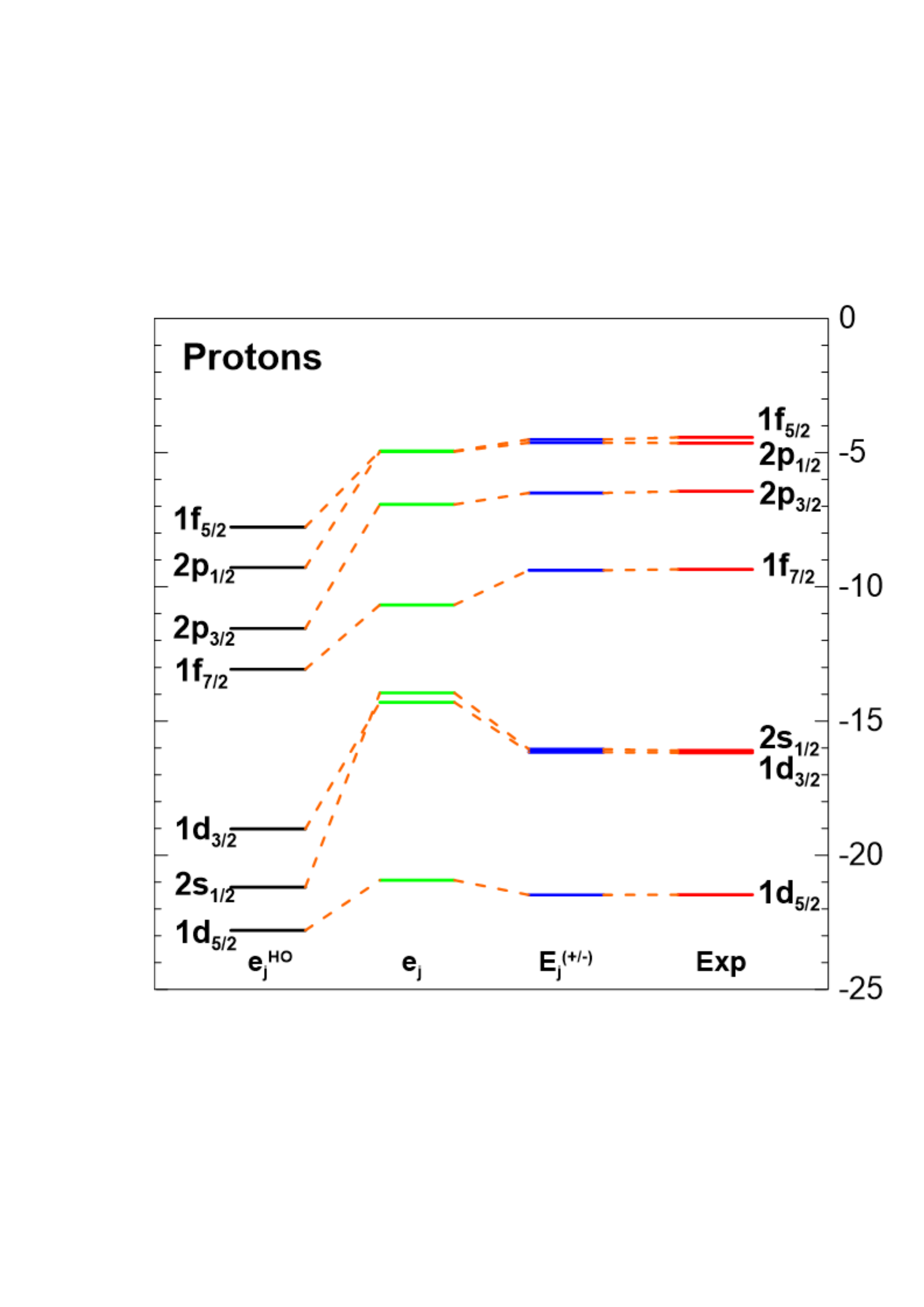}
\end{tabular}
\end{center}
\vspace{-3.cm}
\caption{(Color online) Mean field energies (in units of MeV)
for neutrons (left panel) and for protons (right panel).
In both {cases} are shown: (i)  harmonic oscillator energies
$e^{HO}_j$, (ii)  adjusted single particle energies $e_j$,
(iii)  BCS energies relative to the Fermi
level ($E_{j}^{(\pm)}$),
and (iv) experimental energies ($e^{exp}_j$).}
\label{F2}\end{figure*}
%

\section{Numerical results and discussion}
The residual interaction is described by the
$\delta$-force (in units of MeV$\cdot$fm$^{3}$)
\be
V=-4\pi({\it v}^sP_{s}+{\it v}^tP_{t})\delta(r),
\label{3.1}\ee
where ${\it v}^s$ and ${\it v}^t$ are the spin-singlet
and spin-triplet parameters.

As usually,  the pairing strengths for protons and
neutrons, ${\it v}^s_{pair}({\rm p})$ and
${\it v}^s_{pair}({\rm n})$,  are obtained from the
fitting of the experimental pairing gaps.

In the numerical evaluations of the matrix elements
$G(npn'p' J)$, $F(npn'p'J)$,
$G(n_1n_2n'_1n'_2J_{n})$, $F(n_1n_2n'_1n'_2J_{n})$,
$G(p_1p_2p'_1p'_2J_{p})$, and $F(p_1p_2p'_1p'_2J_{p})$
of the Hamiltonians $H_{pn}$, $H_{pp}$,  $H_{pn}$,
given by Eqs. \rf{32},  \rf{33} and
\rf{34}, were used the same coupling constants.

To set the coupling constants in the $ph$-channel we use the energy
behavior of the IAS and GTR \cite{Nak82} (see also Refs.~\cite{Cas87,Yos18}),
with the results  (in units of MeV$\cdot$ fm$^3$): i)  $v^s_{ph}=27$ and
$v^t_{ph}=64$  for $^{48}$Ca, and ii)   $v^s_{ph}=55$ and
$v^t_{ph}=92$ for all nuclei.

For the coupling constants $v^s_{pp}$ and $v^t_{pp}$ within
the $pp$-channel,
we use values close to those obtained in Ref.~\cite{Fer17} as a result
of the partial restoration of the spin-isospin SU(4) symmetry (PSU4SR).
More precisely,  this procedure yields $s_{sym}\cong t_{sym}\cong  1$
for the ratios
\be
s=\frac{{\it v}_{pp}^s} {{ \overline v}^s_{pair}},
\hspace{1cm}t=\frac{{\it v}_{pp}^t} {{ \overline v}^s_{pair}},
\label{3.2}\ee
where
${\overline v}^s_{pair}=({{\it v}^s_{pair}({\rm p})+{\it v}^s_{pair}({\rm n})})/2$.
\footnote{
Within PSU4SR $s$ and $t$ are determined from the condition that the
strengths $S^+_F \equiv S^{{\{- 1\}}}_0$ and
$S^+_{GT}\equiv S^{{\{- 1\}}}_1$ become minimal (see \cite[Fig. 1]{Fer17}).
}

Moreover, instead of using the bare value $g_{\sss A}=1.27$ for the
axial-vector coupling constant \cite{Ber12},  we use an effective
value $g_{\sss A}=1.0$
\footnote{This quenching is frequently attributed to the $\Delta$-hole
polarization effect on the axial-vector coupling constant \cite{Cas87}.
Recently has been presented an explanation of  the quenching of $g_{\sss A}$
within the context of  effective field theories \cite{Gys19}.}.
Still  smaller values for $g_{\sss A}$ have been used in the
literature~\cite{Suh17}.

\subsection{ Single particle space}
The DBD$^-$ $^{48}$Ca $\go ^{48}$Ti  is a rather unique case,
since  $^{48}$Ca is a double closed nuclei,
and we can make use of the experimental spe $e^{exp}_j$. All they
were taken from the binding and excitation energies,  weighted with
spectroscopic factors, of odd-mass neighboring nuclei:
$^{47}$Ca and $^{49}$Ca for neutrons, and $^{47}$K and $^{49}$Sc for protons.
They  are listed in Fig. \ref{F2} and are those from~\cite[Table II]{Sch07},
except for the proton $f_{5/2}$ spe,  which is estimated from the
proton $f_{5/2}-f_{7/2}$ splitting given in Ref.~\cite{Kor14}.
We need this level to saturate both  the SCE and DCE sum rules.
Once this has been done the spe $e^{exp}_j$ have been used
in two different ways:
\bnu
\item
The following steps are done in handling the BCS
equations~\cite{Con84,Bar95} :\\
a) The BCS energies  relative to the Fermi level $ \lambda$,
\br
E^{(\pm)}_j= \pm E_j + \lambda,
\label{3.3}
\er
are introduced, where the positive (negative) sign is adopted
if the corresponding single-particle state is a particle (hole)-state.
\\
b) It is  assumed that neutron and proton Fermi levels $\lambda_n$,
and  $\lambda_p$ lay between $j_n=2p_{3/2}-1f_{7/2}$
and  $j_p=1f_{7/2}-2s_{3/2}$ states respectively,
and that all states above $\lambda$ are  pure quasi-particle excitations
$E^{(+)}_{j}$ and  all states  below $\lambda$ are pure quasi-hole
excitations $E^{(-)}_{j}$.
\\
c) Starting from a set of harmonic oscillator energies $e_j^{HO}$,
the energies $E^{(\pm)}_j$ are adjusted to the experimental spectra $e^{exp}_j$
by means of a $\chi^2$ search varying the strengths $v_{{s} }^{{pair}}$
and the bare spe $e_j$ which that appear in the BCS gap equations \rf{24}.

All this procedure is illustrated in Fig.~\ref{F2}.
\item
For the sake of completeness the pairing parameters
${\it v}^s_{pair}({\rm p})$ and ${\it v}^s_{pair}({\rm n})$ were
fixed in the standard manner~\cite{Sam10}. That is, by fitting
the experimental pairing gaps to the calculated pairing
gaps $\Delta_j$, given by~\cite[Eq. (2.96)]{Boh69}, with $j=1f_{7/2}$
for neutrons and $j=2s_{1/2}$ for protons.
\enu
The most relevant difference between the spe $e^{exp}_j$ and $e_j$
is the disappearance of the energy gap between the holes and the
particles in the last case. The resulting parameters $v_{{s} }^{{pair}}$,
and $\lambda$ are given in the Table~\ref{T1} for the two sets of
spe $e_j$ and $e_j^{exp}$. The quasiparticle energies $E_{j}^{(\pm)}$
with $e_j^{exp}$ are obviously slightly  different from those shown
in Fig.~\ref{F2}.
\begin{figure*}[th]
\hspace{-2cm}
  \begin{minipage}{0.35\textwidth}
    \includegraphics[scale=0.35]{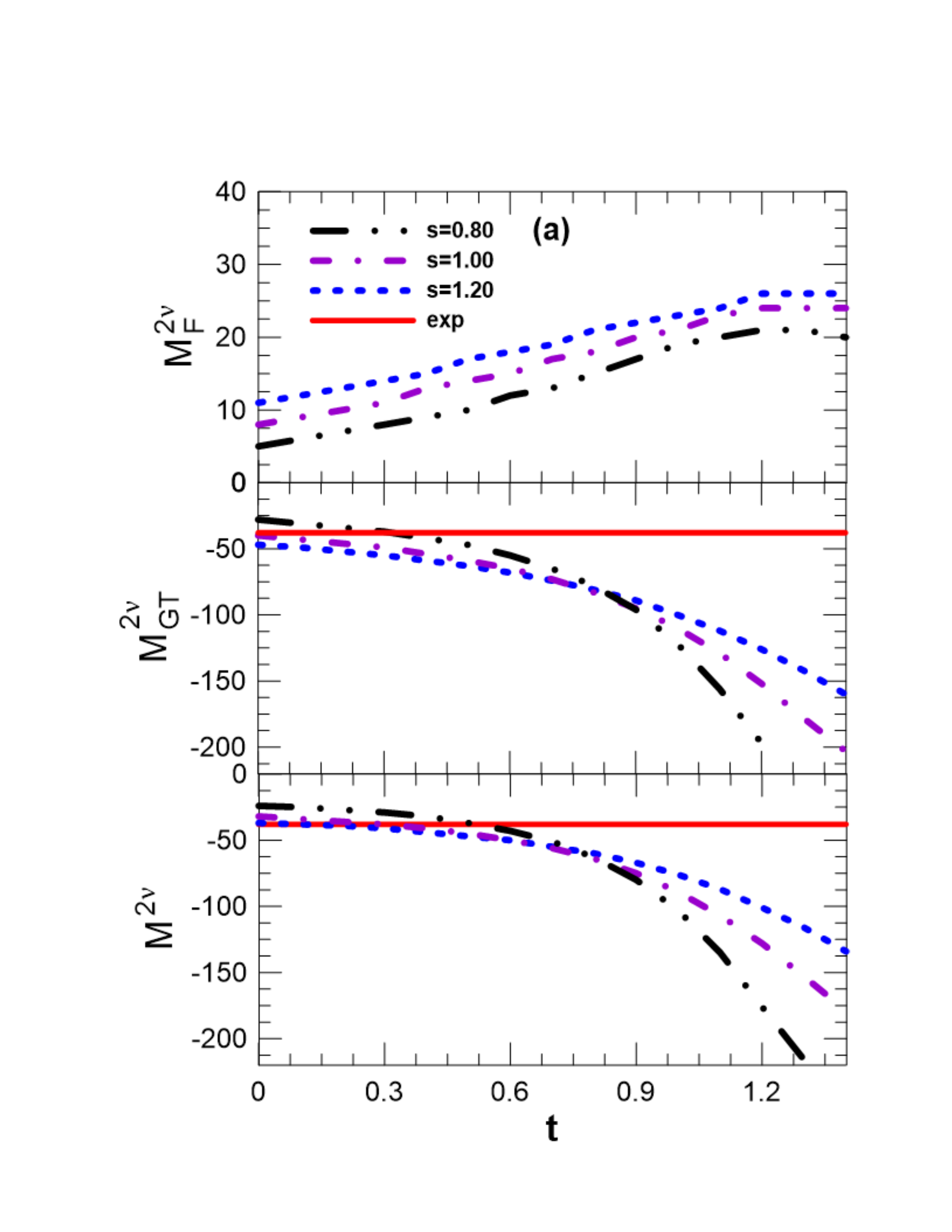}
  \end{minipage}
  \begin{minipage}{0.35\textwidth}
    \includegraphics[scale=0.35]{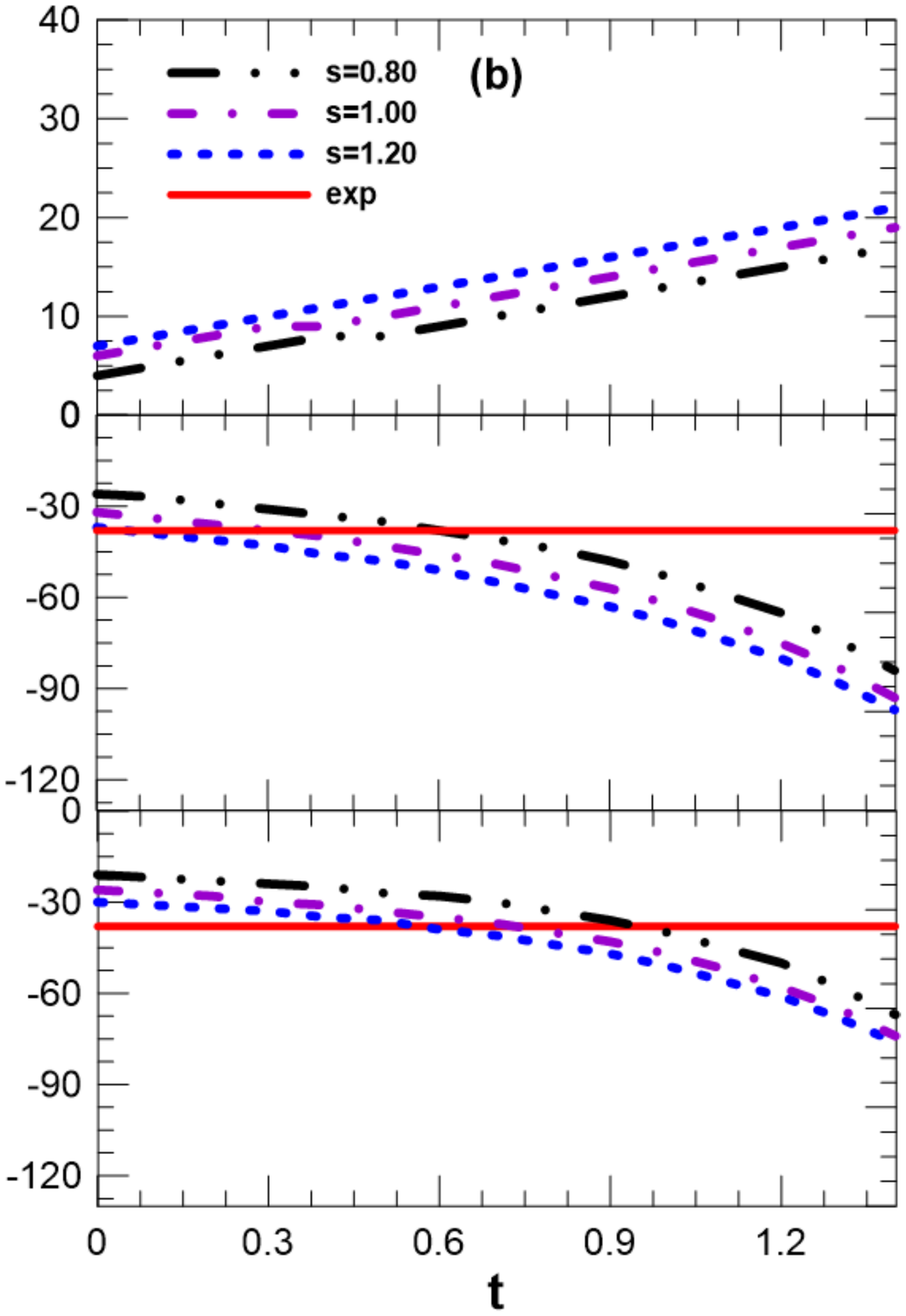}
  \end{minipage}
  \begin{minipage}{0.35\textwidth}
    \includegraphics[scale=0.35]{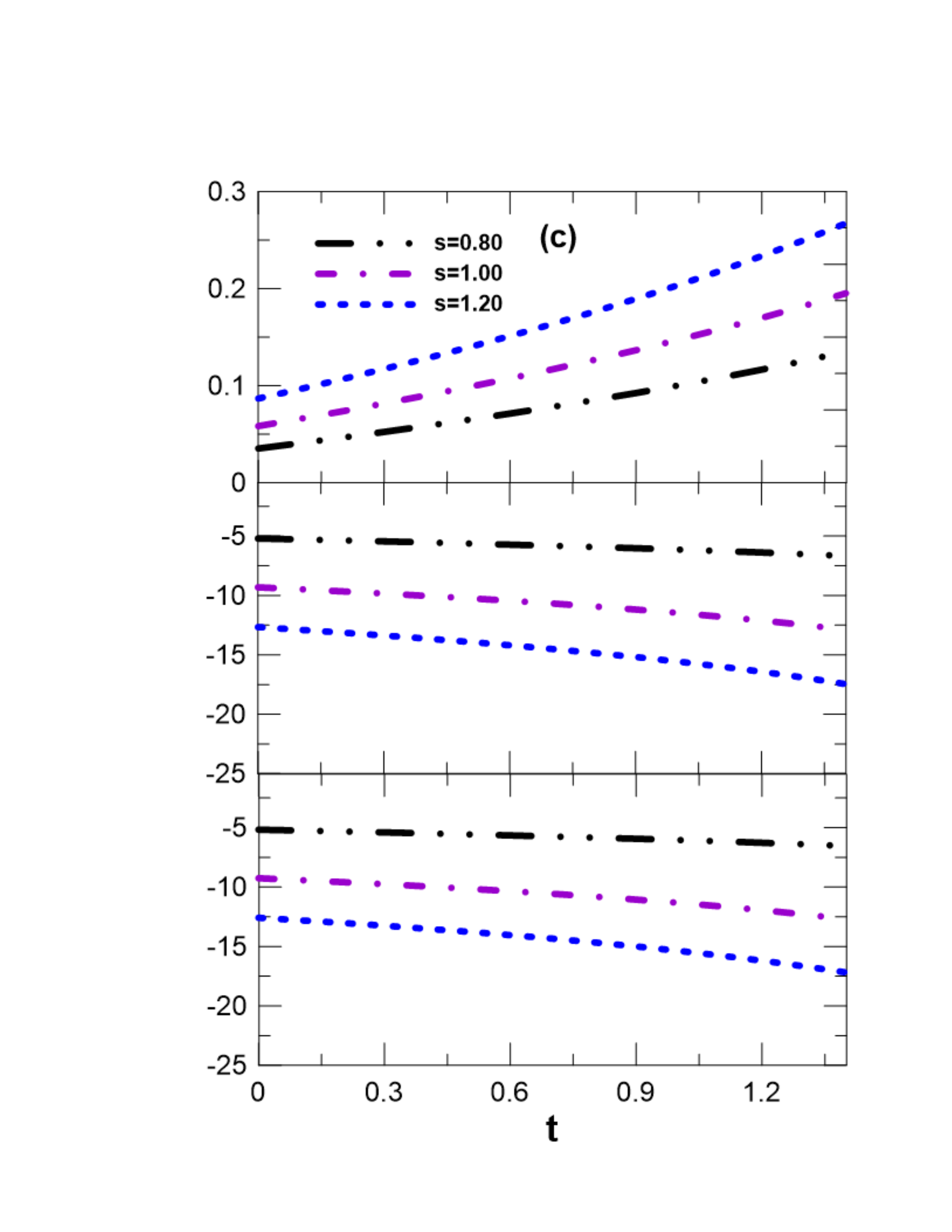}
  \end{minipage}
\vspace{-0.7cm}
\caption{Calculated NME $M^{2\nu}_{F}$, $M^{2\nu}_{GT}$ and $M^{2\nu}$ for the
ground  $0^+$ state (in natural units $\times 10^{-3}$),  as a function of
the $pp$ parameters $t$ and  $s$ for  $^{48}$Ti for:
a) $^{48}$Ca with spe $e^{exp}_j$,  b) $^{48}$Ca with spe $e_j$,
and c) $^{96}$Ru. The experimental value for $^{48}$Ti
$|M^{2\nu}(0^+_1)|=(38\pm 3)\times 10^{-3} $ is also indicated by the red
line, whose width represents the experimental error.}
\label{F3}
\end{figure*}

\begin{table}[th]
\caption
{Results for the BCS coupling constants and Fermi levels.
All notation is explained in the text. The $\lambda$'s
are given in units of MeV, and the couplings $v_{{s}}^{{pair}}$
is in units of MeV$\cdot$fm$^3$.}
\begin{center}
\label{T1}
\begin{tabular}{c|c|cc|cc}
\hline\hline
Nuclei                     & spe         & $v_s^{pair}(n)$ & $v_s^{pair}(p)$ & $\lambda_n$ & $\lambda_p$ \\\hline
\multirow{2}{*}{$^{48}$Ca} & $e^{exp}_j$ & 31.45         & 34.77         & -6.587      & -13.000     \\
                           & $e_j$      & 25.20         & 28.35         & -7.091      & -12.702     \\\hline
$^{96}$Ru                  &      & 33.20         & 38.91         & -8.412      & -5.663 \\\hline\hline
\end{tabular}
\end{center}
\end{table}

In the $^{96}$Ru nucleus, the neutron and proton shells are both open
and its energy spectra is clearly rotational. This fact gives rise to a strong
interplay between collective and single-particle degrees of freedom in the
low energy spectra  of the neighboring odd-mass nuclei $^{95}$Ru, $^{97}$Ru,
$^{95}$Tc, and $^{97}$Ru.
For instance, it is very likely that the ground state $5/2^+$ in $^{95}$Ru is
a consequence of the $j-1$ anomaly in the three neutron cluster
$(1g_{7/2})^3$~\cite{Alm73,Paa73}. Such a complex nuclear structure inhibits us
to determine the appropriate spe spectra and the pairing interaction strengths
from  experimental energy spectra, as it was done in the case of $^{48}$Ca.

\begin{table}[th]
\caption{Neutron and proton spe for $^{96}$Ru, which were obtained
in the way  explained in the text, together with
the resulting quasiparticle energies \rf{3.3}.
All notation is explained in the text. The energies
are given in units of MeV, and the couplings
in units of MeV$\cdot$fm$^3$.}
\begin{center}
\label{T2}
\newcommand{\cc}[1]{\multicolumn{1}{c}{#1}}
\renewcommand{\tabcolsep}{0.5pc}
\begin{tabular}{c| r r |  r r}\hline
\multirow{2}{*}{level} & \multicolumn{2}{c|}{Neutrons} &\multicolumn{2}{c}{Protons}\\\cline{2-5}
           &$e_j$      &$E_{j}^{(\pm)}$ & $e_j$    & $E_{j}^{(\pm)}$  \\\hline
 $3s_{1/2}$&$  -5.396$ &$ -5.1817$     & $  4.916$&  $ 5.006$        \\
 $2d_{3/2}$&$  -5.236$ &$ -4.9834$     & $  4.792$&  $ 4.897$        \\
 $1g_{7/2}$&$  -6.878$ &$ -6.5267$     & $  2.308$&  $ 2.502$        \\
 $2d_{5/2}$&$  -7.401$ &$-10.0514$     & $  2.493$&  $ 2.627$        \\
 $1g_{9/2}$&$ -14.401$ &$-14.5004$     & $ -5.424$&  $-7.448$        \\
 $2p_{1/2}$&$ -17.802$ &$-17.9127$     & $ -7.718$&  $-8.478$        \\
 $2p_{3/2}$&$ -19.389$ &$-19.4838$     & $ -9.295$&  $-9.773$        \\
\hline\hline
\end{tabular}
\end{center}\end{table}

We used instead the spe provided by N. Paar~\cite{Paa19}, which
were calculated in the relativistic Hartree-Bogoliubov model,
as outlined in Ref.~\cite{Paa08}. They are shown in Table~\ref{T2},
together with the resulting quasiparticle energies~\rf{3.3},
which were obtained following procedure 2 in the case
of $^{48}$Ca. This implies to fit the calculated pairing
gaps $\Delta_j$, with $j=1g_{7/2}$ for neutrons and $j=1g_{9/2}$
for protons, to the experimental ones.
The similarity between the spe $e_j$  and the quasiparticle
energies $E_{j}^{(\pm)}$  is remarkable. The corresponding
pairing parameters ${\it v}_s ^{pair}$ and  chemical
potentials $\lambda$ are listed in Table \ref{T1}.

It is important to note that in $^{48}$Ca it is $\lambda_n >\lambda_p$
while in $^{76}$Ru this difference is $\lambda_n < \lambda_p$.
We will soon see that this fact is decisive with respect
to the $Q$-values of DBD. More precisely, this will explain
why DBD$^-$ occurs in $^{48}$Ca and DBD$^+$ occurs in $^{76}$Ru.

Our method of calculation is similar in several aspects to that used
in the SM in the Refs.~\cite{Sag16, Aue18}. In fact, the illustration of their
calculations, made in Fig. 1 of that reference, is also valid in our case.
The biggest difference between the two models, in addition to the residual
interactions that were used, arises from the size of the configuration
spaces of the final states $\J_f^+$. We have 664 $0^+$ states
and 2.470  $2^+$ states, while Auerbach and Bui Minh Loc~\cite{Aue18}
have, in their evaluation of double charge-exchange GT
strength $^{48}$Ca $\go ^{48}$Ti, the quantity  of 14.177
and 61.953 final states in the $fp$-space, respectively. In the
case of the $^{96}$Ru $\go ^{96}$Mo decay, we also have 664 $0^+$ states,
but 2.583  $2^+$ states.
\footnote{ See also our Fig. \ref{F1},  where
the  difference with the standard pn-QRPA calculation of the NME
is illustrated graphically.}

\subsection{Nuclear Matrix Elements}
\begin{table}[h]
\caption{
Calculated and measured NME (in natural units $\times 10^{-3}$)
for  $2\nu$-DBDs of $^{48}$Ca ($M^{2\nu}(\J^+_f)\equiv M^{2\nu^-}(\J^+_f)$) and
$^{96}$Ru ($M^{2\nu}(\J^+_f)\equiv M^{2\nu^+}(\J^+_f)$) to
ground state $0^+_1$, and first excited $0^+_2$ and $2^+_1$ states in
final nuclei $^{48}$Ti and $^{96}$Mo, respectively.
Results from Refs.\cite{Rad07,Hor07} for $^{48}$Ca and
from Refs.\cite{Suh12,Hir94,Rum98,Rai06,Bar15} for $^{96}$Ru,
as well the experimental value for   $^{48}$Ca \cite{Bar11},
are also shown in same units.}\label{T3}
\centering
\vspace{-0.05cm}
\begin{tabular}{c|c|c|c|c|c}\hline\hline
\multicolumn{6}{c}{$^{48}$Ca}   \\\hline\hline
par                           & $M^{2\nu}_F(0^+_1)$ & $M^{2\nu}_{GT}(0^+_1)$ & $|M^{2\nu}(0^+_1)|$ & $|M^{2\nu}(0^+_2)|$ & $|M^{2\nu}(2^+_1)|$ \\\hline
$e_j$&&&&&\\
T1                            & 11                  & -44                    & 33                  & 18                  & 0.98                \\
T2                            & 13                  & -53                    & 40                  & 22                  & 1.08                \\
T3                            & 15                  & -62                    & 47                  & 15                  & 0.69                \\\hline
$e_j^{exp}$&&&&&\\
T1                            & 15                  & -78                    & 63                  & 5.0                 & 0.53                \\
T2                            & 19                  & -122                   & 103                 & 23                  & 0.60                \\
T3                            & 21                  & -110                   & 89                  & 34                  & 0.33                \\\hline
\cite{Rad07}&&&22&&120  \\\hline
 \cite{Hor07}&&&28&26&1.62  \\\hline
Exp.&&&$38\pm 3$&& \\\hline
\hline
\multicolumn{6}{c}{$^{96}$Ru}                                                                                                                  \\\hline\hline
T1                            & 0.08                & -5.9                   & 5.8                 & 55.6                & 0.55                \\
T2                            & 0.10                & -6.1                   & 6.0                 & 63.6                & 0.68                \\
T3                            & 0.15                & -11.5                  & 11.3                & 26.8                & 0.63                \\\hline
\cite{Suh12}                   &                     &                        & 415-1437            & 492-1554            & 0.1-8.4             \\
\cite{Hir94} &                     &                        & 251                 &                     &                     \\
\cite{Rum98} &                     &                        & 101                 &                     &                     \\
\cite{Rai06} &                     &                        & 54                  &                     &                     \\
\cite{Bar15} & -0                  & 2170                   & 2170                & 50                  &  \\\hline\hline
\end{tabular}
\end{table}

We calculate simultaneously the NMEs $M^{2\nu}_F(0^+_f)$,
$M^{2\nu}_{GT}(0^+_f)$,  $M^{2\nu}(0^+_f)$, and  $M^{2\nu}_{GT}(2^+_f)$
for all above mentioned $\J^+_f$ final states
with the  following three sets of $pp$ parameters:
\br
T1: &&s=0.80,\hspace{.5cm} t=0.80,\nn\\
T2: &&s=0.80,\hspace{.5cm} t=1.00,\nn\\
T3: &&s=1.00,\hspace{.5cm} t=1.00.
\label{3.4} \er

In the upper part of  Table \ref{T3} are shown the results
for the $^{48}$Ca $\go ^{48}$Ti decays  to the ground state $0^+_1$,
and the first excited $0^+_2$ and $2^+_1$ states in $^{48}$Ti nucleus,
for the two sets of  spe $e_j$  listed in Table \ref{T1}.
The agreement between the calculated and measured results
for $M^{2\nu}(0^+_1)$ can be considered satisfactory
(in particular with the spe $e_j$) in view of the fact that all
the nuclear parameters in the $pp$ and $ph$-channels are so to say fixed,
both for identical particles and for different
particles. We hope that in the next future the NMEs $M^{2\nu}(0^+_2)$
and  $M^{2\nu}_{GT}(2^+_1)$ will also be measured.
For both $0^+$ levels the F-NME is relatively small compared
to the GT one, but in no way it can be neglected. As seen in the
Eq.~\rf{1}, these two NMEs always interfere destructively.
\footnote{It is interesting to note that $M_F$ is often omitted
in the calculations, simply invoking isospin conservation.}

The NME $|M^{2\nu}(0^+_1)| $ in  $^{48}$Ca has been calculated
many times, but there are only very few theoretical studies
of $M^{2\nu}(0^+_2)$\ and $M^{2\nu}_{GT}(2^+_1)$. As far as we
know, the first one has been evaluated only in Ref.~\cite{Hor07}
and the second one in Refs.\cite{Rad07,Hor07}
(see also Ref. \cite{Hor13}). These results, as well as those for
NME $ |M^{2\nu}(0^+_1)|$,  are confronted with our results in
Table~\ref{T3}. It should be pointed out that in the just mentioned
studies have not been considered the contributions of F-NMEs
$M^{2\nu}_F(0^+_{\a=1,2})$. Therefore, strictly speaking their
results for $ |M^{2\nu}(0^+_{\a=1,2})|$ should be compared with ours
$|M^{2\nu}_{GT}(0^+_{\a=1,2})|$.

In the lower part of Table \ref{T3} are shown the NMEs for
the $^{96}$Ru $\go ^{96}$Mo decay to final states $\J^\pi=0^+_1,0^+_2$,
and $2^+_1$.
The results of previous calculations~\cite{Suh12,Hir94,Rum98,Rai06,Bar15}
are also shown. It is noticeable that the differences between our three
calculations are much smaller than the differences with all the other works.

The strong dependence of the NME within the pn-QRPA model with respect
to the isoscalar $pp$ parameter $t$ is well known and is often discussed.
Therefore, it could be interesting to analyze that dependence in the
current model. This is done in Fig.~\ref {F3} showing the NMEs
$M^{2\nu}_{F}$, $M^{2\nu}_{GT}$  and $M^{2\nu}$  for the ground
$0^+$ state in $^{48}$Ca  and  $^{96}$Ru, as a function of
the $pp$ parameters  $t$ and $s$. The experimental value
of $|M^{2\nu}(0^+_1)| $ is also drawn. It can be concluded that within
the present model such dependence is only moderate.
The same statement is valid also for all remaining $2\nu$ NMEs.

It is also well known that the relatively small values of the NME in
the pn-QRPA model come from the destructive interference between forward
and backward going contributions~\cite{Fer17}. That is, trough  the
ground state correlations (GSC). The quenching mechanism is different
in the current model, and it is the  consequence of the interplay
between seniority-zero and seniority-four configurations in the
final states.
For example, in the case of  $^{48}$Ca,  within the space $e_j$ and
with force parameters T1, the NME  $M^{2\nu}$ for the three lowest  $0^+$
states are:  $ -0.033,    -0.018, -0.043$.
While, when only the seniority-zero configurations are considered,
one gets: $-0.074,-0.028, -0.090$ respectively, which confirms
the above statement.

\subsection {Half-lives}
\begin{table}[th]
\begin{center}
\caption{Calculated half-lives $\tau_{2\nu}^{2\b^-}(\J^+_f=0^+_1,0^+_2,2^+_1)$
in units of $yr$ for the $2\nu$-DBD $^{48}$Ca $\go ^{48}$Ti,
with spe $e_j$ the  $pp$ parameters set and $ g_{\sss A}=1$ are
shown and confronted with previous calculations and  experiments. In Eq.
\rf{47} are used the $G^\a_{2\nu}(\J^+_f)$ factors from Ref.~\cite{Kot12}
for the levels $0^+_1(=1.56 \times 10^{-17}$ yr$^{-1} )$ and
$0^+_2(=3.63 \times 10^{-22}$ yr$^{-1} )$, and from Ref.~\cite{Doi83}
for $2^+_1(=4.41 \times 10^{-18}$ yr$^{-1} )$.}
\label{T4}
\begin{tabular}{c|c|c|c}
 \hline\hline
                 & $ 0^+_1$           &$0^+_2$  &$2^+_1$ \\\hline
T1          &$5.91\times 10^{19} $&$ 8.51 \times 10^{24}  $&$2.36 \times 10^{23}$ \\\hline
T2          &$4.02\times 10^{19} $&$ 5.70 \times 10^{24}  $&$1.94 \times 10^{23}$ \\\hline
T3          &$2.91\times 10^{19} $&$ 1.23 \times 10^{25}  $&$4.76 \times 10^{23}$ \\\hline
Ref. \cite{Rad07}&$              $&$                      $&$1.72 \times 10^{24}$ \\\hline
Ref. \cite{Hor07}&$3.3\times 10^{19}$&$  $&$ 8.5\times 10^{23} $\\\hline
Exp  \cite{Bar11}&$ \left(4.4^{+0.6}_{-0.5}\right)\times 10^{19}$&& \\\hline \hline
\end{tabular}
\end{center}
\end{table}

\begin{table}[th]
\begin{center}
\caption{Calculated half-lives $\tau_{2\nu^\pm}^{\a}(0^+_1)$
in units of $yr$ for the $2\nu$-DBD $^{96}$Ru $\go ^{96}$Mo,
with the $pp$ parameter set  T1 and $ g_{\sss A}=1$
are shown and confronted with previous calculations and
experiments. In Eq. \rf{47} are used the $G^\a_{2\nu}(\J^+_f)$
factors from Ref.~\cite{Doi92} for the channels
$2\b^+(= 1.080\x10^{-26}$ yr$^{-1} )$, $\b^+e (= 0.454\x10^{-21}$ yr$^{-1} )$,
and $ee (=2.740\x10^{-21}$ yr$^{-1} )$.}
\label{T5}
\begin{tabular}{c|c|c|c|c|c}
\hline \hline
$\a$            &         present         &    \cite{Hir94}     & \cite{Rum98}        &     \cite{Rai06}      &\cite{Bar15}  \\\hline \hline
$2\b^+$         &    $2.8\times 10^{30}$  & $5.8\times 10^{26}$ &                     & $3.485\times 10^{28}$ & $1.22\times 10^{16}$ \\\hline
$\b^+e$         &    $6.5\times 10^{25}$  & $1.2\times 10^{22}$ & $8.6\times 10^{22}$ & $9.100\times 10^{23}$ & $3.78\times 10^{20}$\\\hline
$ee$            &    $1.1\times 10^{25}$  & $2.1\times 10^{21}$ & $1.4\times 10^{22}$ & $1.628\times 10^{23}$ & $3.11\times 10^{17}$\\\hline \hline
\end{tabular}
\end{center}
\end{table}

The half-lives are evaluated trivially from \rf{47} using the NME
and leptonic kinematic factors $G^\alpha_{2\nu}(\J^+_f)$.
Despite this, we present some of them only for the sake of completeness.

In Table \ref{T4} are compared our results for the
half-lives $\tau_{2\nu^-}^{2\b^-}(\J^+_f=0^+_1,0^+_2,2^+_1)$ in $^{48}$Ca
evaluated with spe $e_j$ $pp$ parameters set with the previous ones.

The major difference appears for the $2^+_1$  level.

The same is done in Table \ref{T5} for the half-lives $\tau_{2\nu^+}^{\a}(0^+_1)$
in  $^{96}$Ru for different channels  $\a=2\b^+$, $\b^+ e$, and $ee$.
The experimental limits are: $\tau_{2\nu^+}^{2\b^+}(0^+_1)\ge 1.4 \times 10^{20}$ yr
and  $\tau_{2\nu^+}^{\b^+e}(0^+_1)\ge 0.8\times 10^{20}$ yr~\cite{Bel13}.

\begin{figure*}[th]
\hspace{-2cm}
  \begin{minipage}{0.35\textwidth}
    \includegraphics[scale=0.40]{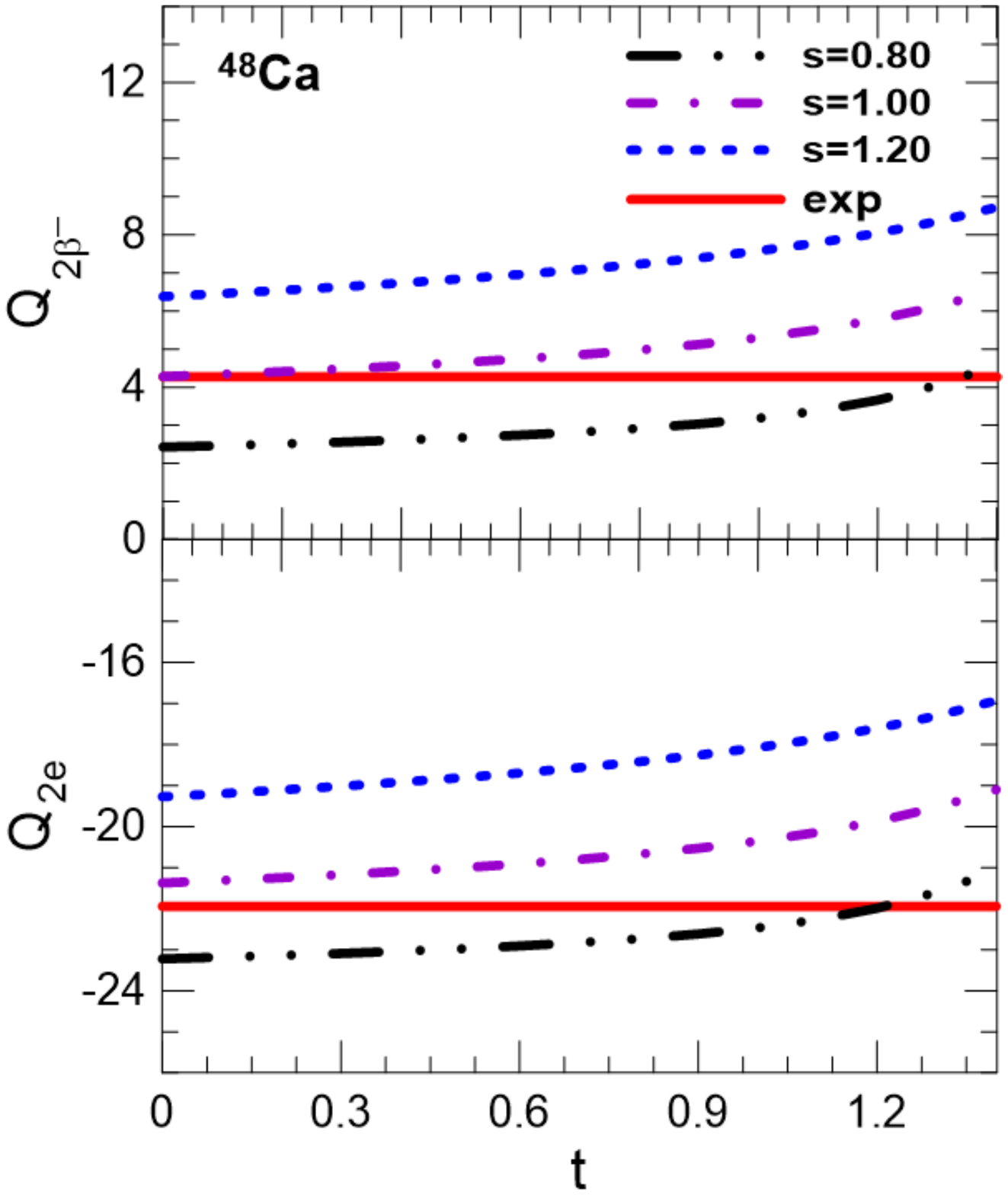}
  \end{minipage}
  \hspace{0.5cm}
  \begin{minipage}{0.35\textwidth}
      \includegraphics[scale=0.40]{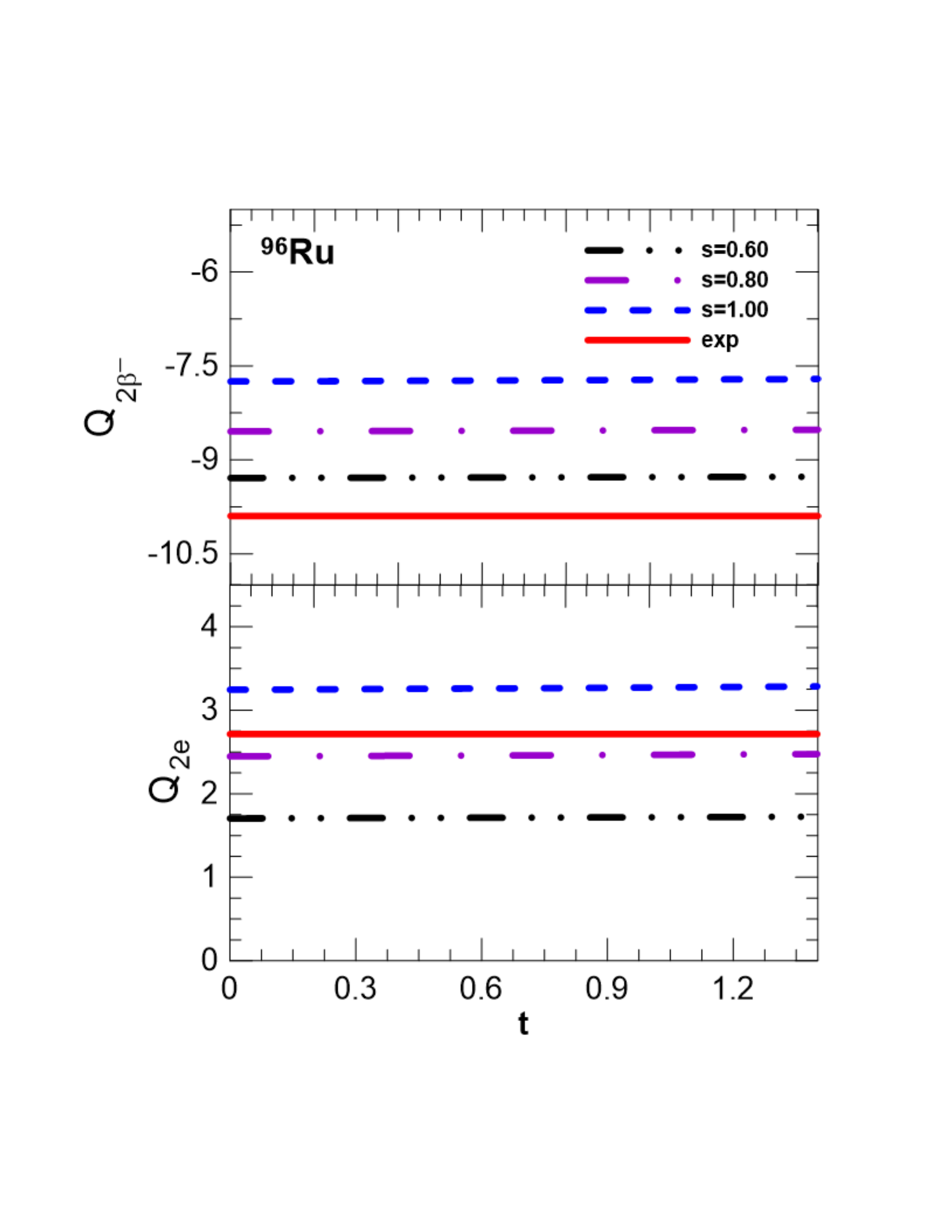}
  \end{minipage}
\vspace{-2.cm}
\caption{Calculated $Q_{2\b^-}$ and $Q_{2e}$ values in
$^{48}$Ca within the spe space $e^{exp}_j$ (left panel)
and  $^{96}$Ru (right panel), as a function of  $pp$ parameters $t$
and  $s$. The experimental $Q$-values are also shown.}\label{F4}
\end{figure*}

\begin{figure}[h]
\begin{center}
\vspace{-1.cm}
    \includegraphics[scale=0.45]{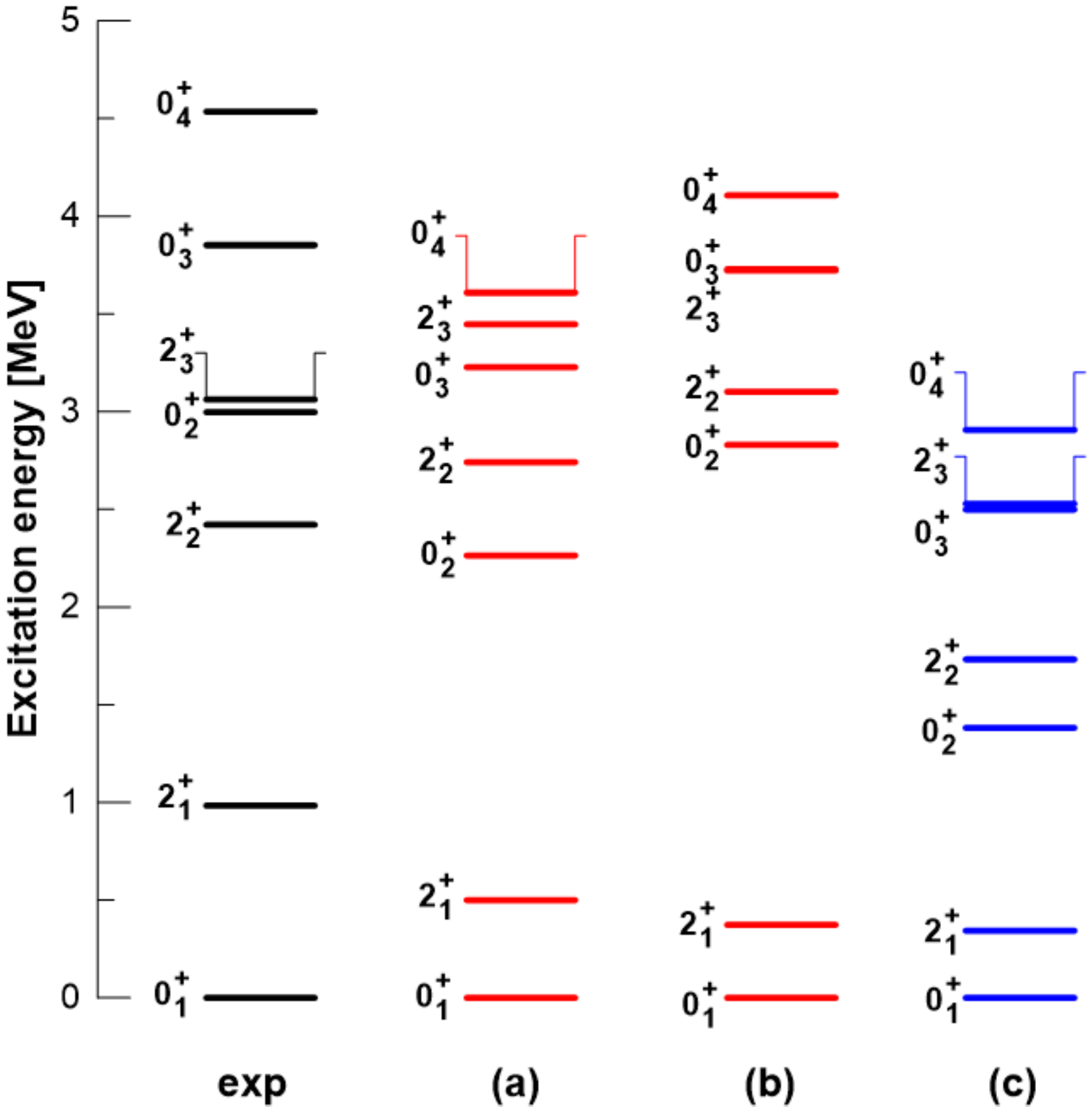}
\vspace{-1.9cm}
\caption{Measured excitation energies in $^{48}$Ti are compared with
the calculations: (a) and  (b)  with the spe $e^{exp}_j$
and $pp$ parametrization T1 and T3 respectively,
and (c)  with the spe $e_j$ and $pp$ parametrization T1.
}\label{F5}\end{center}
\end{figure}

\subsection {$Q$-values and Energy Spectra}
\begin{table}[th]
\begin{center}
\caption{Calculated $Q_{\b\b\-}$ and $Q_{2e}$ values (in units of MeV)
with the three sets of $pp$ parameters T1, T2 and T3 and for the
DCE processes: a)  $^{48}$Ca $\go ^{48}$Ti and  $^{48}$Ca $\go ^{48}$Ar,
and b) $^{96}$Ru $\go ^{96}$Mo and  $^{96}$Ru $\go ^{96}$Pl,
are confronted with the experimental ones.}
\label{T6}
\begin{tabular}{c|c|cc}
 \hline\hline
 \multicolumn{4}{c}{$^{48}$Ca} \\\hline\hline
                   &  par &$ Q_{\b\b\-}  $&$  Q_{2e}   $\\ \hline
\hline
\multirow{3}{4em}{$e_j^{exp}$}& T1&$  2.919    $&$   -22.734   $\\
                              & T2&$  3.185    $&$   -22.468   $\\
                              & T3&$  5.300    $&$   -20.353   $\\ \hline
\multirow{3}{4em}{$e_j$}      & T1&$  5.151    $&$   -17. 295   $\\
                              & T2&$  5.232    $&$   -17.214   $\\
                              & T3&$  6.710    $&$   -15.736   $\\ \hline
      Exp &&$  4.268   $&$  -21.943    $\\ \hline \hline
 \multicolumn{4}{c}{$^{96}$Ru} \\\hline\hline
\multirow{3}{4em}{$e_j$}      & T1&$  -8.532   $&$   2.460   $\\
    & T2&$  -8.527   $&$   2.465   $\\
                              & T3&$  -7.721   $&$   3.271   $\\ \hline
     Exp &&$  -9.896   $&$  2.714    $\\ \hline\hline
 \end{tabular}
\end{center}
\end{table}

Before starting with the discussion of $Q$-values, it is convenient
to remember that a physical phenomenon is allowed only when this quantity
is positive.

In Table \ref{T6} are confronted the experimental data with our results for
the $Q_{\b\b\-}$ and $Q_{2e}$ values in the DCE processes:
a)  $^{48}$Ca $\go ^{48}$Ti and  $^{48}$Ca $\go ^{48}$Ar,
and b) $^{96}$Ru $\go ^{96}$Mo and  $^{96}$Ru $\go ^{96}$Pl.
One sees that the model is capable of reproducing not only the signs
of the $Q$-values, but also their magnitudes, without having to modify the
parameters of the model. This is very comforting! In addition, it seems that
the model ``knows" what type of DCE decay can occur in a given nucleus.

The nature of $Q$-value is dominantly determined by the proton and neutron
pairing mean fields, as seen from \rf{41} or, more precisely, from the
relation $Q_{2e}-Q_{2\b^-}=4(\lambda_n-\lambda_p)$. The dependence on
the residual interaction is rather weak and takes place through the ground
state energy $\w_{0_1^+}$ in residual nuclei, as $Q_{2e}+Q_{2\b^-}=-2\w_{0_1^+}$.
More details on how the $Q$-values depend on the $pp$ coupling constants
are shown in the Fig. \ref{F4}.

As we stated before, in the same way that the pn-QRPA model~\cite{Hal67}
predicts identical energy spectra for odd-odd nuclei  $(A,Z\pm 1)$,
the present model predicts identical excitation energies in even-even
nuclei  $(A,Z\pm 2)$.  This is obviously not realistic due to the
large neutron excess.

It is pertinent to mention here that the use of particle-number-projection
can become very important when working with the BCS mean-field~\cite{Krm93}.
Without a doubt, through this method different energy spectra are obtained
in nuclei for which the number of protons is different.

But, despite the above mentioned handicap, the calculated excitation energies
of the  $0^+$ and $2^+$ states in $^ {48}$Ti are consistent with data, as
shown in Fig.~\ref{F5}, where we give the calculated spectra for the two
spe spaces and the $pp$ parameterizations T1 and T3.
Because of the size of the  $Q_{2\b^-}$-value ($=  4.268 $ MeV),
the $2\b^-$-decays are energetically possible for all states, except
for $0^+_4$. We have evaluated the NMEs for all these states, but we do not
consider  necessary to present them here. It should also be said that we
have not found in the literature any detailed calculation of the $^ {48}$Ti
low energy spectrum to compare with ours. Finally, the $^{96}$Mo energy spectrum
is not well explained by the current model and, therefore, will not be
discussed here.

\subsection {Double-Charge-Exchange Strengths and their Sum Rules}

In Table \ref{T7} are displayed the results for the DCE
transition  strengths $ S^{\{\pm 2 \}}_{J\J}$ given by \rf{15},
both for Fermi ($J=0;\J=0$) and Gamow-Teller ($J=1;\J=0,2$).
The corresponding sum rules  $S^{\{2\}}_{J\J}$ calculated from \rf{16}
are also shown,  and confronted with the predicted sum rules
${\sf S}^{\{2\}}_{J\J}$ given by \rf{17},  \rf{18}, and \rf{19}.

In addition, to know the locations of the DCE resonances,
we have evaluated the centroid energy~\rf{46}, and to get
an idea of the magnitudes of the DBD, the strengths
$B^{\{+2\}}_{J\J_1}$ ($ B^{\{-2\}}_{J\J_1}$) going to the
levels $0_1^+$ and $2_1^+$ in  final  $^{48}$Ti ($^{96}$Mo)
nucleus are explicitly given.

All calculations related to DCE transition strengths were performed
for the three sets of $pp$ parameters \rf{3.4}, finding that all
produce identical results. We have also found that, at least in
the case of $^{48}$Ca, there is a certain dependence of the
results with respect to the spe spaces used.

In the last three rows of the upper part of the Table \ref{T7},
are given the results derived for $^{48}$Ca by other authors
within the SM for the $pf$-space.
\footnote{Note that the present  calculations were done in
a single-particle space consisting of the $2p-1f -2s-1d$
shells for both protons and neutrons.}
Namely, by i) Sagawa and Uesaka~\cite{Sag16} with GXFF1A interaction,
and by ii) Auerbach and Minh Loc \cite{Aue18}, and Shimizu, Men\'endez
and Yako~\cite{Shi18}, both with KB3G interaction. In fact, the values
of strengths $S^{\{+ 2 \}}_{1\J=0,2}$ attributed to the last
authors~\cite{Shi18} have been extracted from their Fig. 1.

\begin{table}[th]
\caption{Results with the $pp$ parametrization $T1$ for:
i)  Fermi ($J=0;\J=0$) and Gamow-Teller ($J=1;\J=0,2$) DCE
transition strengths $ S^{\{\pm 2 \}}_{J\J}$ given by \rf{15},
ii) the corresponding sum rules  $S^{\{2\}}_{J\J}$ calculated
from\rf{16}, iii) the predicted  sum rules $ {\sf S}^{\{2\}}_{J\J}$
given by \rf{17},  \rf{18}, and \rf{19}, iv) the energy centroid~\rf{49},
and v) the transition strengths $ B^{\{2\}}_{J\J_1} (\equiv B^{\{+2\}}_{J\J_1}$) for $^{48}$Ca
and $ B^{\{2\}}_{J\J_1} (\equiv B^{\{-2\}}_{J\J_1}$) for $^{96}$Ru going to the
levels $0_1^+$ and $2_1^+$. The SM results from previous works~\cite{Sag16, Aue18, Shi18}
for $^{48}$Ca are also shown. The meaning of the inequalities is explain in the text.
}
\label{T7}
\begin{tabular}{c|c|ccccccc}
\hline\hline
\multicolumn{9}{c}{$^{48}$Ca} \\\hline\hline
\multirow{2}{*}{Ref.}  & \multirow{2}{*}{$ J\J $} & \multirow{2}{*}{$ S^{\{-2\}}_{J\J}$}& \multirow{2}{*}{$ S^{\{+2\}}_{J\J}$ }&
\multirow{2}{*}{$ S^{\{2\}}_{J\J}$} & \multirow{2}{*}{${\sf S}^{\{2\}}_{J\J}$} & \multirow{2}{*}{${\bar E_{00}}^{\{-2\}}$} &
\multirow{2}{*}{${\bar E_{00}}^{\{+2\}}$} & $ B^{\{2\}}_{J\J_1}$\\
&&&&&&&&$\times 10^{-3}$\\ \hline
            &$00$&$  140.1  $&$ 0.94  $&$  139.1  $&$ 112         $&$   20.5  $&$   -     $&$ $ 77$  $\\
$e_j^{exp}$ &$10$&$  162.2  $&$ 2.0   $&$  160.2  $&$ \le  175.9  $&$ 12.2    $&$   15.9  $&$470$\\
           &$12$&$  716.1  $&$ 9.10  $&$ 707.0   $&$ \ge 640.0   $&$ 13.2    $&$    16.5 $&$51 $\\
 \hline
            &$00$&$  157.1  $&$ 2.38  $&$  139.1  $&$ 112         $&$   21.6  $&$ 22.7    $&$45$\\
$e_j$       &$10$&$  189.8  $&$5.87   $&$  160.2  $&$\le 183.9    $&$   14.1  $&$   19.7  $&$163$\\
           &$12$&$  858.5  $&$  26.1 $&$ 832.4   $&$\ge 752.5    $&$   14.8  $&$   19.7  $&$52$\\
\hline
\cite {Sag16}&$10$&$ -  $&$-    $&$  135.5$&$\le 144.0  $&$   -  $&$  -  $&-\\
             &$12$&$ -  $&$  -  $&$ 501.2 $&$\ge 480.0  $&$   -  $&$  -  $&-\\\hline
\cite {Aue18} &$10$&$  131.8$&$- $&$-$&$\le 144.0 $&$  21.9  $&$  -  $&$ 0.24  $\\\hline
 \cite {Shi18}&$10$&$  126.3  $&$- $&$ -$&$- $&$  -  $&$  -  $&$   -$\\
             &$12$&$ 511.0  $&$  -  $&$ -$&$-$&$   -  $&$  - $&$   -  $\\
\hline
\multicolumn{9}{c}{$^{96}$Ru}                                                                                                                  \\\hline\hline
            &$00$&$  128.0  $&$  0.1 $&$  127.9 $&$ 112         $&$ 21.4   $&$  20.4 $&   $0.0016$ \\
$e_j$  &$10$&$  221.1  $&$ 12.0 $&$  209.1 $&$ \le 222.7   $&$ 23.1   $&$  13.1 $&   $ 16$\\
            &$12$&$  981.4  $&$ 49.4 $&$  932.0 $&$ \ge 873.4   $&$ 23.1   $&$  12.3 $&   $ 5.7$\\
\hline \hline
\end{tabular}
\end{table}

Several observations are in order regarding the results
shown in Table \ref{T7}:
\bnu
\item The strengths $ S^{\{- 2\}}_{J\J}$ are always
small in comparison with  the strengths $ S^{\{+2\}}_{J\J}$
and as a  consequence $ S^{\{+ 2\}}_{J\J}\cong{\sf S}^{\{ 2\}}_{J\J}$.
This is clearly due to the relatively large neutron excess.

\item Although small, the strengths $ S^{\{- 2\}}_{J\J}$ are significant
in relation to the DBD. They are proportionally higher in $^{96}$Ru,
which decays by $\b^+\b^+$, than in $^{48}$Ca, which decays by $\b^-\b^-$.

\item The  F strengths $S^{\{ 2\}}_{00}$ deviate quite significantly
from the sum rule strengths ${\sf S}^{\{ 2\}}_{00}$; $24 \% $ and  $40\%$,
respectively,  within the spe spaces $e_j^{exp}$ and $e_j$ in $^{48}$Ca,
and $14\%$  in $^{96}$Ru. A possible explanation for these differences
is given in the appendix.
\footnote{In this case it could be interesting to analyze if the problem
can be solved by particle-number projection \cite{Krm93} . }
\item Terms proportional to $ C $ in the GT sum rules  \rf{18}, \rf{19},
and \rf{20}   {are} not included in the calculations,
and this is the reason why  have to be fulfilled the conditions
\br
S^{\{2\}}_{10}\le {\sf S}^{\{2\}}_{10},
\nn\\
 S^{\{2\}}_{12}\ge {\sf S}^{\{2\}}_{12}.
\label{3.5} \er
In fact,  they are nicely satisfied in all numerical calculations
presented in  Table \ref {T7}.

\item All $S^{\{\mp 2\}}_{J\J}$ strengths depend quite significantly
on the  spe, but very weakly on the residual interaction.
This is the reason why we only show the results for the parametrization T1.

\item The same situation applies to the predicted sum
rules ${\sf S}^{\{ 2\}}_{1\J}$, due to their dependence on the term
$S^{\{+1\}}_1$ (in Eqs.~\rf{18}, \rf{19} and \rf{20}), which in turn
depends on the spe used in the calculations.

\item The terms proportional to $C$ are omitted in \rf{18}, \rf{19}
in Refs. ~\cite{Sag16, Aue18},  and this is the reason why
their predicted sum rules ${\sf S}^{\{ 2\}}_{1\J}$ are smaller
than ours.

\item Since the values for $ S^{\{+2\}}_{1\J}$ are not explicitly
given in Shimuzu~\etal\cite {Shi18},  we have derive them from their Fig. 1b.
They are consistent with the values of
${\sf S}^{\{ 2\}}_{1\J}$ presented in Refs.~\cite {Sag16, Aue18}.

\item Our GT strengths are always larger than those in the SM calculations.
Also our average energies  ${\bar E_{1\J}}^{\{+2\}}$ are significantly
smaller that those presented in Table II in Ref.~\cite {Aue18}, and those
shown in Fig. (1b) by  Shimuzu \etal~\cite {Shi18}. It is difficult to
discern whether this is due to the deficiency of our model, or the
difference in the size of single-particle spaces. We are inclined to
think that our results are correct, since otherwise it would
be very difficult to satisfy the second condition in Eq.~\rf{3.5}.

\item The size of $ B^{\{\pm 2\}}_{J\J_1}$,
shown in the last column of Table \ref {T7}, and when compared
with $ S^{\{\pm 2\}}_{J\J}$, give us an idea on the smallness
of the NMEs.
\enu
\begin{figure}[th]
\begin{center}
    \includegraphics[scale=0.45]{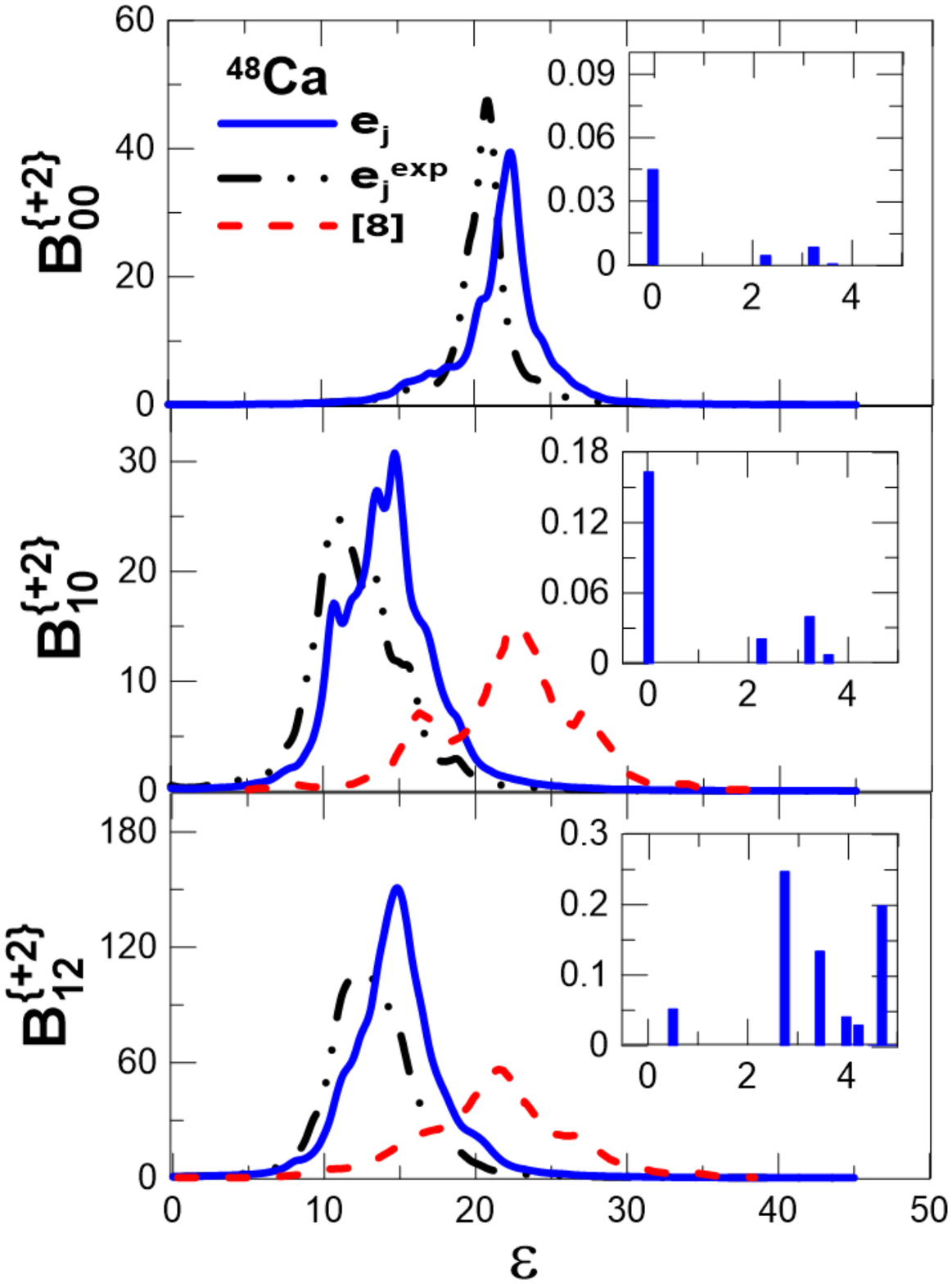}
\vspace{-1.2cm}
\caption{ DCE strength distributions $B^{\{+ 2\}}_{J\J}$ for
the transition $^ {48}$Ca $\go^ {48}$Ti
with  spe  $e^{exp}_j$, and $e^{exp}_j$, and $pp$ strengths T1,
as a function of the excitation energy $\E$ in $^ {48}$Ti.
The $B^{\{+ 2\}}_{J\J}$ are dimensionless, and the energies
are in MeV. The SM  results, obtained by Shimizu \etal~\cite{Shi18}
with the KB3G interaction, are also shown.}
\label{F6}
\end{center}
\end{figure}

\begin{figure}[h]
\begin{center}
    \includegraphics[scale=0.45]{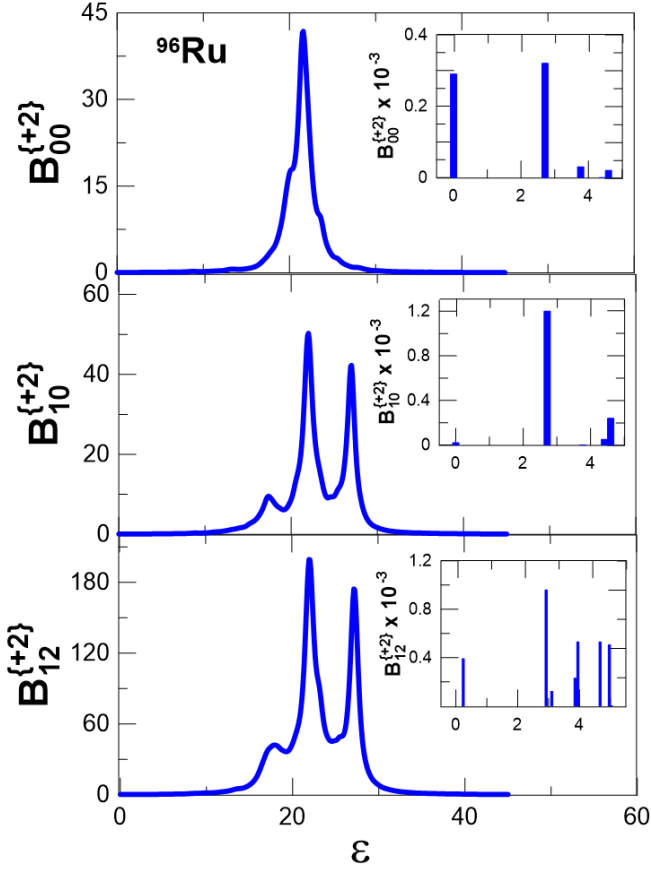}
\vspace{-0.9cm}
\caption{DCE strength distributions $B^{\{+ 2\}}_{J\J}$
for the transition $^ {96}$Ru $\go^ {96}$Pd.
The $B^{\{- 2\}}_{J\J}$  are dimensionless, and the energies are in MeV.}
\label{F7}
\end{center}
\end{figure}

\begin{figure}[h]
\begin{center}
    \includegraphics[scale=0.45]{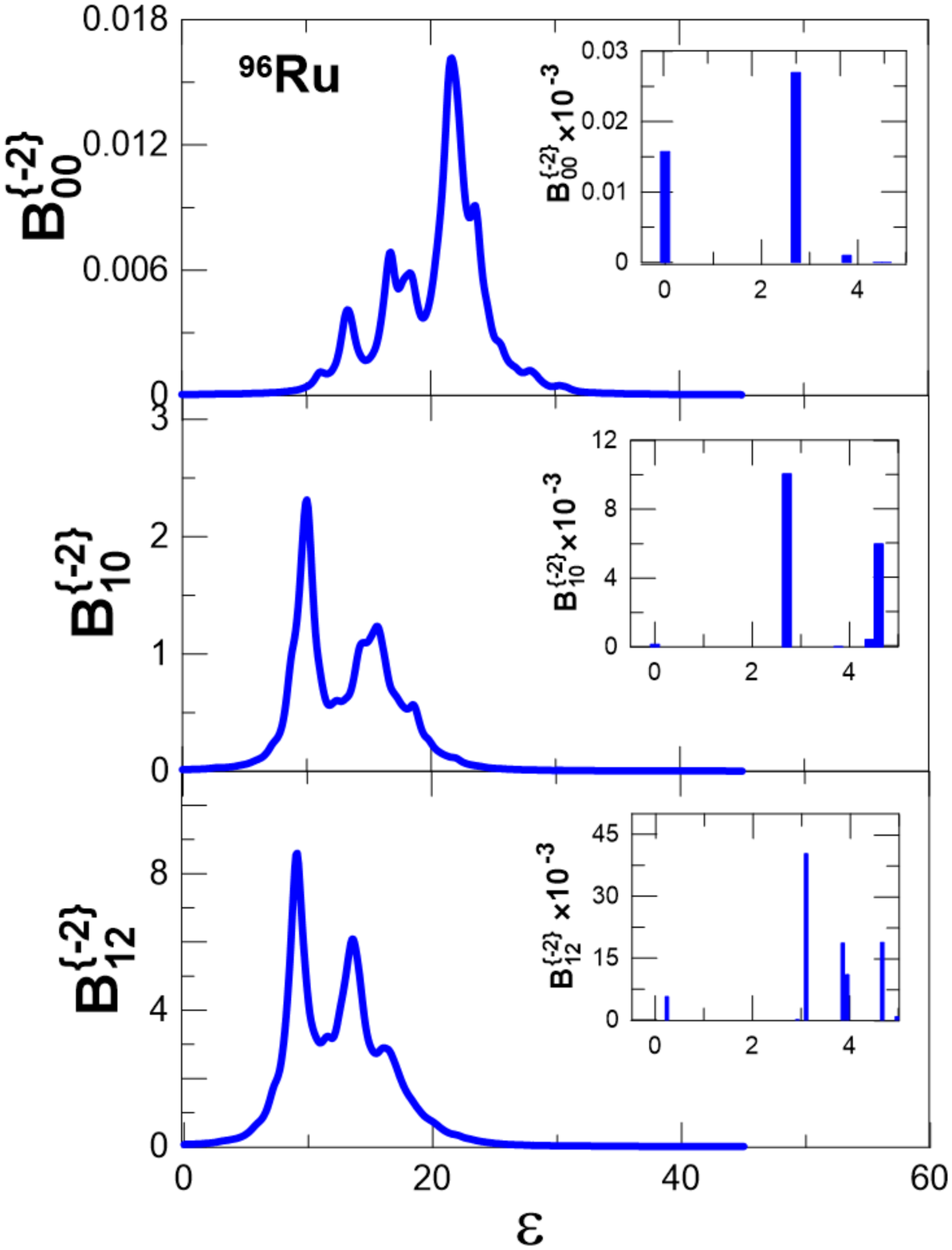}
\vspace{-0.9cm}
\caption{DCE strength distributions $B^{\{- 2\}}_{J\J}$ for the transition $^{96}$Ru $\go ^{96}$Mo.
The $B^{\{- 2\}}_{J\J}$   are dimensionless, and the energies are in MeV.}
\label{F8}
\end{center}
\end{figure}

\subsection{Spectral Distributions of Double Charge Exchange Strengths }

The  DCE strength distributions  $B^{\{\pm 2\}}_{J\J_f}$, which are
of interest here, are drawn in Figs.~\ref{F6}-\ref{F8}
as a function of the excitation energy $\E$ in final nuclei.  We have
found that they depend only moderately on the spe spaces, and even less
on the $pp$ parameters. To simulate the experimental energy resolution,
they were smeared out with Lorentzians of 1 MeV width. Moreover,
these figures contain inserts which show the corresponding  strengths in the
low-lying states of final nuclei.

In the upper panel of Fig. \ref{F6} are  shown the F distributions
{$B^{\{+2\}}_{00}$}  in $^{48}$Ti, exhibiting  at around $ 22 $ MeV a fairly
narrow double giant F resonance, usually called double isobaric analog state (DIAS).
In the middle and lower panels of this figure are shown the GT
distributions $B^{\{+2\}}_{10}$ and {$B^{\{+2\}}_{12}$}, respectively,
in the same final nucleus, which also exhibit resonant like structure.
These double GT giant resonances (DGTGR) are much wider than the DIAS
and centered around 13 and 14 MeV, respectively.
In the  KB3G SM calculations  of Shimizu \etal~\cite{Shi18}, which
are also shown in Fig.~\ref{F6}, these resonances appear at around 20 MeV.

In Figs. \ref{F7} and  \ref{F8}  are presented  analogous results for the
{$B^{\{+ 2\}}_{J\J}$} and {$B^{\{-2\}}_{J\J}$} densities in  $^{96}$Pd and
$^ {96}$Mo final nuclei, respectively.
Both are shown because here we are interested in the $DBD^+$, where
the low-energy behavior of $B^{\{-2\}}_{J\J}$ densities is relevant.

As seen in  Fig.~\ref{F7}, the DIAS in $^{96}$Pd is located at
around $21$ MeV, while both $\J=0^+$ and $\J=2^+$ DGTGR are at about $23$ MeV.
These resonances are not directly related to the  DBD of $^{96}$Ru, but
their locations in $^{96}$Pd can be searched through heavy ion reactions.

The smallness of {$B^{\{-2\}}_{00}$} and its energy distribution,
shown in  Fig.~\ref{F8}, are fully consistent with the small
value of $M^{2\nu}_F(0^+_1)$ in Table \ref{T3} and of {$B^{\{-2\}}_{00_1}$}
in Table \ref{T7}. Moreover, the distributions of the {$B^{\{-2\}}_{1\J=0,2}$}
clearly indicate that the DBD$^+$ of $^{96}$Ru will be very slow.

\subsection{Comparison  between  ground state  $2\nu$-DBD NME
and DCE strengths}

In the so-called closure approximation, the sum of the intermediate
states $J_\a$ in \rf{1} and \rf{9} are taken by closure, after
replacing $E_{J_\a }^{\{\mp 1\}}$ in \rf{3} by some average
$\bar E_{J}^{\{\mp 1\}}$~\cite{Doi83}.

\begin{figure*}[th]
\vspace{-2.cm}
\begin{center}
\begin{tabular}{cc}
\includegraphics[width=8cm]{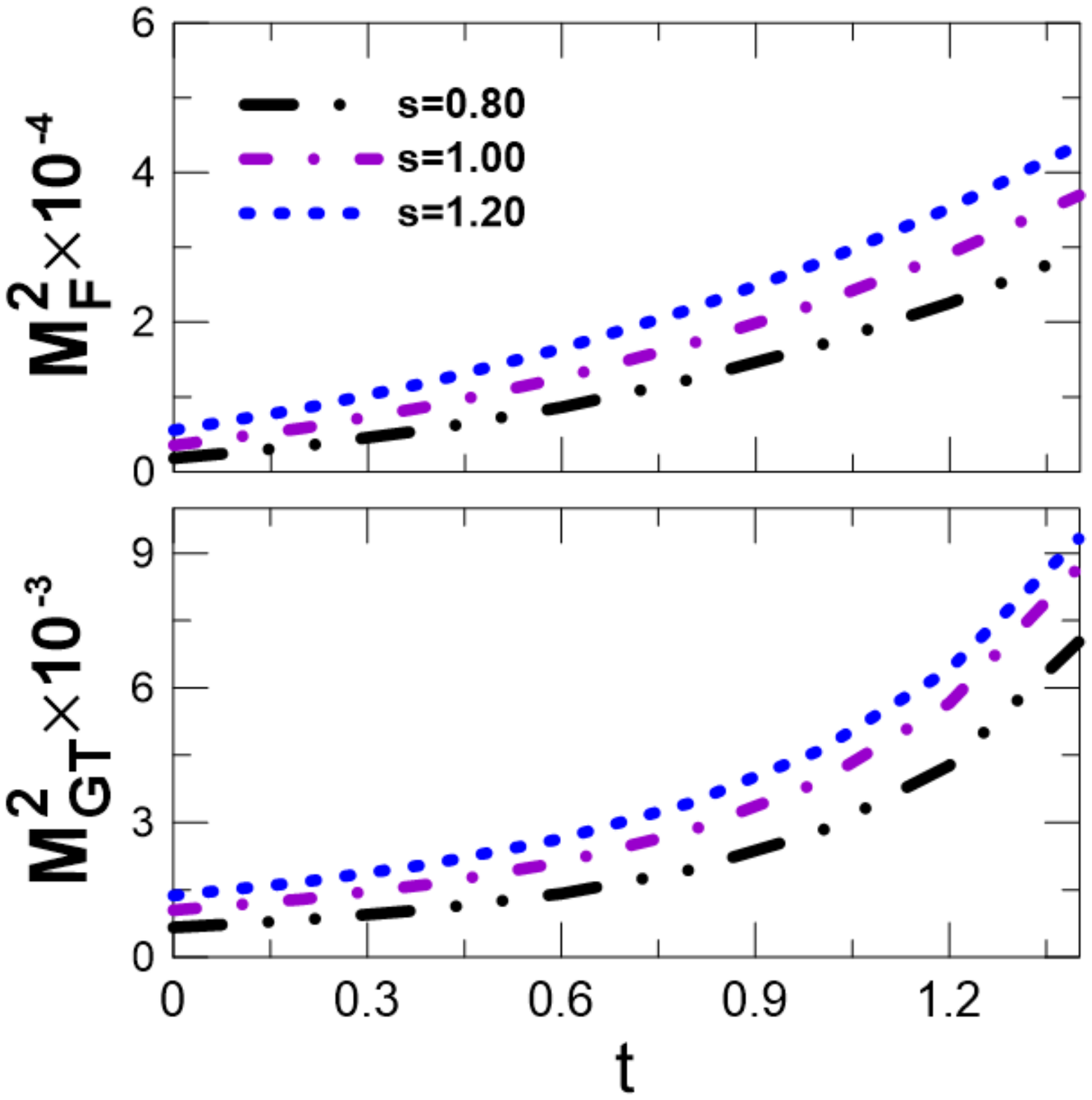}
&
\includegraphics[width=8cm]{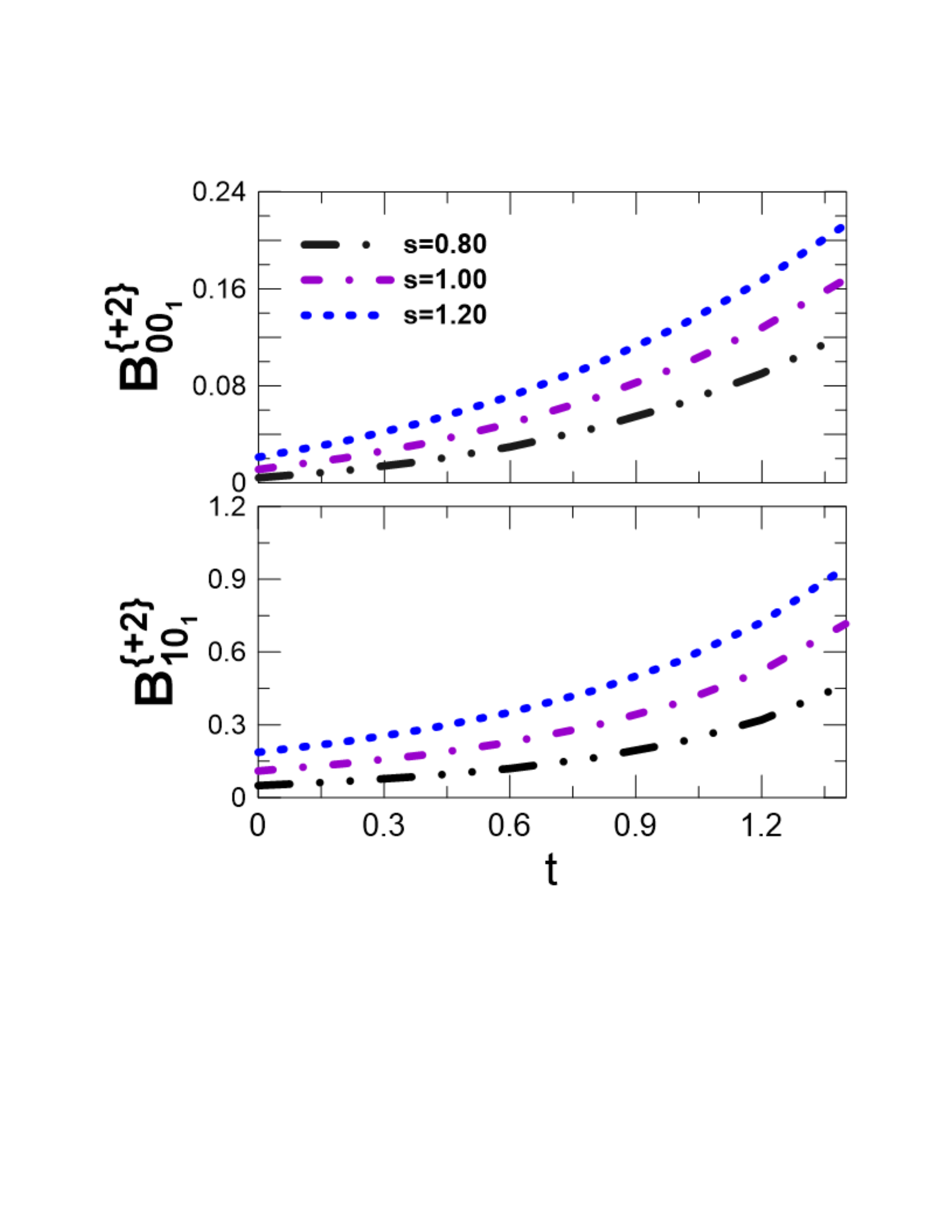}
\end{tabular}
\end{center}
\vspace{-3.cm}
\caption{(Color online) Comparison of  $|M^{2\nu}_F(0^+_1)|^2$
and $|M^{2\nu}_{GT}(0^+_1)|^2$ (both in natural units)
with $B^{\{+2\}}_{00_1}$  and $B^{\{+2\}}_{10_1}$ (dimensionless),
respectively, in   $^{48}$Ca for the spe $e_j$.}
\label{F9}\end{figure*}

Thus, except for the constant energy denominator, the ground state
DCE densities  $B^{\{\pm 2\}}_{J0_1}$  are the closure approximations
of the squares of the NMEs.  In view of this, to know how reasonable
the closure approximation is, it may be interesting to compare the
behaviors of these two quantities as a function of the $pp$ parameters.
As an example, in Fig. \ref{F9} are compared the squares of the
NMEs  $M^{2\nu}_F(0^+_1)$ and  $M^{2\nu}_{GT}(0^+_1)$ with $B^{\{+2\}}_{00_1}$,
for $^{48}$Ca with the spe $e_j$. The squares of the NMEs are in
natural units, while the strengths are dimensionless.
The proportionality between these two observables suggests that
the closure approximation in the case of $^{48}$Ca is reasonable.
However, there is no guarantee that this result
will be valid in general.
\section {Final Remarks}

We have developed  a nuclear structure model that involves
(pn,2p2n)-QTDA excitations on the BCS mean-field, which
is capable of simultaneously describing the DBD and the
DCE transition strengths. So far, this has been done only
in the context of SM, where these two problems are generally
treated separately, although it is well known that they are
intimately related to each other.  This is the case,
for instance, of $^{48}$Ca,  where the DBD$^-$s are described
in Refs.~\cite{Rad07,Hor07}, while the transition strength
distributions $B^{\{+ 2\}}_{J\J}$ and the corresponding total
strengths $S^{\{+2\}}_{J\J}$ were evaluated in Refs.~\cite {Sag16,Aue18, Shi18} .

The  (pn,2p2n)-QTDA model has additional advantages over the standard
pn-QRPA model. Namely:

\begin{enumerate}
\item Together with the NMEs of the ground state, the NMEs of
all the excited states $0^+$ and $2^+$  are calculated simultaneously.
To do the same in the pn-QRPA model, it is necessary to resort to
supplementary calculations through several charge-conserving QRPAs,
thus introducing several new model parameters.
\item It allows the evaluation of the $Q$-values for DBDs, which plays
a very important role in this type of processes.
\end{enumerate}

The proposed  model can be viewed as a natural extension to DCE
processes of the pn-QRPA model, originally  developed  by HS to
describe the SCE processes~\cite{Hal67}. The first does not
include the GSC like the second.  But this is not a serious
inconvenience since, as we have discussed above, the quenching mechanism
is now different.

Our next aim is to evaluate and discuss the $0\nu$-NMEs~\rf{12} making use
of the  replacement \rf{11} in our previous work \cite{Fer17}.
One expects that the relationship between DCE nuclear reactions and DBD
will be more clearly visible at $ 0\nu $ than at $ 2\nu $ reactions,
due to a lower dependence on the NME of their energy denominators
in the first case.

During the development of the present study, Santopinto \etal \cite{San18},
based on a previous work of Bertulani~\cite{Ber93}, have reported that,
in the low-momentum-transfer limit, the  heavy ion
$^{40}$Ca($^{18}$O, $^{18}$Ne)$^{40}$Ar cross section behaves as
\brn
\frac{d\sigma}{d\Omega}&\sim\left|\frac{\M^{DGT}_{T\go T'}
\M^{DGT}_{P\go P'}}{{\bar E_{P}}^{GT}+{\bar E_{T}}^{GT}}+\frac{\M^{DF}_{T\go T'}
\M^{DF}_{P\go P'}}{{\bar E_{P}}^{F}+{\bar E_{T}}^{F}}\right|^2,
\label{4.1}
\ern
where $P$ and $T$ stand for projectile and target nuclei respectively.

The correspondence with our notation is:
\\
1)  For the matrix elements
\footnote{Except for the coupling constants $c_{GT}$
and $c_{GT}$ in  Eqs. (11) and (12) respectively.}
\brn
\M^{DGT}_{P\go P'}\go B^{\{+ 2\}}_{10_1},
\hspace{0.5cm} \M^{DF}_{P\go P'}\go B^{\{+ 2\}}_{00_1},
\nn\\
\M^{DGT}_{T\go T'}\go B^{\{- 2\}}_{10_1},
\hspace{0.5cm} \M^{DF}_{T\go T'}\go B^{\{+ 2\}}_{00_1}.
\label{4.2}
\ern
2) For  the energies (see the denominator in  \rf{2})
\brn
E_{P}^{GT}&\go &E_{1_\a }^{\{+ 1\}}-E_{0^+}^{\{0\}},
\hspace{0.2cm}E_{P}^{F}\go E_{0_\a }^{\{+ 1\}}-E_{0^+}^{\{0\}},
\nn\\
E_{T}^{GT}&\go &E_{1_\a }^{\{-1\}}-E_{0^+}^{\{0\}},
\hspace{0.2cm}E_{T}^{F}\go E_{0_\a }^{\{- 1\}}-E_{0^+}^{\{0\}}.
\label{4.3}
\ern
Therefore, the present model posses all the necessary
ingredients to evaluate the heavy-ion cross section in the
low-momentum-transfer limit. Of course, now it is necessary
to solve two eigenvalue problems, one for the target nucleus
$^{40}$Ca, and one for the projectile nucleus $^{18}$O.

In summary, we have developed a new model, based on the BCS approach,
to describe the double-charge exchange nuclear phenomena $(A,Z)\rightarrow (A,Z\pm 2)$.
It is  a natural extension  of the  Halbleib and Sorensen~\cite{Hal67}
model, aimed to describe the single-charge exchange processes $(A,Z)\rightarrow (A,Z\pm 1)$.
As an example, detailed  numerical calculations are presented for
the  $(A,Z)\rightarrow (A,Z+ 2)$ process in $^{48}$Ca $\rightarrow ^{48}$Ti
and the $(A,Z)\rightarrow (A,Z- 2)$ process in $^{96}$Ru $\rightarrow ^{96}$Mo,
involving all final $0^+$ states and $2^+$ states.
At the moment we are extending this study in two directions:
\bnu
\item A throughout evaluation of all $2\nu$- DBD$^\pm$,
together with the associate nuclear reaction strengths will performed.
\item The $2\nu$- DBD$^\pm$ formalism developed here will be
extended to the $0\nu$- DBD$^\pm$.
\enu

\begin{acknowledgements}
This study was financed in part by the Coordena\c{c}\~ao de Aperfei\c{c}oamento
de Pessoal de N\'ivel Superior Brasil (CAPES)  Finance Code 001.
A.R.S. acknowledges the financial support of FAPESB
(Funda\c{c}\~ao de Amparo \`a Pesquisa do Estado Bahia)
TERMO DE OUTORGA- PIE0013/2016.
The authors thank the partial support of UESC (PROPP 00220.1300.1832).
We sincerely thank to Wayne Seale for his very careful
and judicious reading of the manuscript.
We  also thank N. Paar for providing us the spe for $^{96}$Ru,
evaluated within the (DD-ME2) model, and to C. Bertulani and C. Barbero
for stimulating comments and discussions.
\end{acknowledgements}

\appendix*
\section{A Toy Model}\label{A}

In order to understand why F DCESR is not completely satisfied
in our model, we resort to a toy model corresponding to the $^{14}$C nucleus
and considering the levels $1s_{1/2}$, $1p_{1/2}$, and $1p_{3/2}$, of which
the three are totally occupied by neutrons, while only the first two are
partially occupied by protons.  From \rf{14} we have
\br
S^{\{- 2\}}_{00}&=&\sum_{ f}B^{\{-2\}}_{00_f}
\nn\\
&=&\sum_{ f}
|\sum_{\a }\Bra{0^+_f}\O^-_0\Ket{0_\a^+}
\Bra{0_\a^+}\O^-_0\ket{0_i^{+}}|^2.
\label{A1}\end{eqnarray}
In the BCS approximation one gets
\brn
B^{\{-2\}}_{00_1}&=&4u_{1p_{1/2}}^4,
\hspace{.1cm}B^{\{-2\}}_{00_2}=4u^2_{1s_{1/2}}u^2_{1p_{1/2}}, \hspace{.1cm}
\nn\\
B^{\{-2\}}_{00_3}&=&12u^2_{1s_{1/2}}u^2_{1p_{1/2}},
\hspace{.1cm}B^{\{-2\}}_{00_4}=4u_{1s_{1/2}}^4,
\ern
and
\br
S^{\{- 2\}}_{00}\equiv S^{2\beta^{-}}_F=4+8u^2_{1s_{1/2}}u^2_{1p_{1/2}},
\label{A2}\er
since $u_{1s_{1/2}}^2+u^2_{1p_{1/2}}=1$.
For instance, with $u^2_{1s_{1/2}}=0.95$ and $u^2_{1p_{1/2}}=0.05$,
one gets $S^{\{- 2\}}_{00}=4.382$, instead of the predicted
value $S^{\{- 2\}}_{00}=4$.
The result \rf{A2} also is valid when the residual interaction
is switched on. This means that the F DCESR
is fully satisfied only in the particle-hole limit, {\it i.e.}
when one of the protons $1s_{1/2}$, $1p_{1/2}$ levels
is totally full or totally empty.

\end{document}